\documentclass[a4paper,11pt]{article}

\pdfoutput=1

\usepackage{jheppub}
\usepackage[utf8]{inputenc}
\usepackage{amsmath, amssymb}
\usepackage{mathrsfs}
\usepackage{float}
\usepackage{color}
\usepackage{xcolor}
\usepackage{multirow}
\usepackage{graphicx,caption,subcaption}
\usepackage{bm}
\usepackage{latexsym}
\usepackage{slashed}
\usepackage[space]{grffile}
\usepackage{enumitem}
\usepackage{gensymb}
\usepackage{textcomp}
\usepackage[title,titletoc]{appendix}
\newcommand{\mrm}[1]{{\mathrm{#1}}}
\newcommand{\trm}[1]{{\textrm{#1}}}
\newcommand{\Ga}{{\Gamma}}
\newcommand{\De}{{\Delta}}

\newcommand{\al}{{\alpha}}
\newcommand{\bt}{{\beta}}
\newcommand{\ga}{{\gamma}}
\newcommand{\de}{{\delta}}

\newcommand{\lm}{{\lambda}}

\newcommand{\bbZ}{{\mathbb{Z}}}
\newcommand{\cL}{{\mathcal{L}}}
\newcommand{\cM}{{\mathcal{M}}}
\newcommand{\cN}{{\mathcal{N}}}
\newcommand{\rb}{{\mathrm{b}}}
\newcommand{\rL}{{\mathrm{L}}}
\newcommand{\rR}{{\mathrm{R}}}
\newcommand{\sfH}{{\mathsf{H}}}
\newcommand{\sfI}{{\mathsf{I}}}
\newcommand{\sfL}{{\mathsf{L}}}

\newcommand{\dg}{{\dagger}}

\newcommand{\ol}{\overline}
\newcommand{\pd}{{\partial}}

\DeclareMathOperator*{\msum}{\mbox{$\sum$}}

\title{Dark Dimension Right-handed Neutrinos Confronted with Long-Baseline Oscillation Experiments}

\author[a]{Ai-Yu Bai}
\author[b]{Auttakit Chatrabhuti}
\author[a]{Yin-Yuan Huang}
\author[b]{Hiroshi Isono}
\author[a]{Jian Tang}

\affiliation[a]{School of Physics, Sun Yat-Sen University, 510275 Guangzhou, China}
\affiliation[b]{High Energy Physics Research Unit, Faculty of Science, Chulalongkorn University,\\ Phayathai Road, Pathumwan, Bangkok 10330, Thailand}

\emailAdd{baiaiy@mail2.sysu.edu.cn, auttakit.c@chula.ac.th, huangyy378@mail2.sysu.edu.cn, hiroshi.i@chula.ac.th, tangjian5@mail.sysu.edu.cn}

\abstract{Right-handed neutrinos are naturally induced by dark extra dimension models and play an essential role in neutrino oscillations. The model parameters can be examined by the long-baseline neutrino oscillation experiments. In this work, we compute the predicted neutrino oscillation spectra within/without extra dimension models and compare them with the experimental data. We find that the neutrino data in the T2K and NOvA experiments are compatible with the standard neutrino oscillation hypothesis. The results set the stringent exclusion limit on the extra dimension model parameters at a high confidence level. 
The derived constraints on dark dimension right-handed neutrinos are complementary to those results from the collider experiments and cosmological observations.}

\begin{document}

\maketitle

\section{Introduction}

Extra dimensions are a crucial component of String Theory and are often considered to be compact to align with the four-dimensional description of our observable universe. There are good physical motivations that their size may be significantly larger than the four dimensional Planck length, accounting for large hierarchies of energy scales in Nature~\cite{Antoniadis:1990ew,Arkani-Hamed:1998jmv,Antoniadis:1998ig}. Recently, the Swampland program, particularly the Distance Conjecture~\cite{Ooguri:2006in, Lust:2019zwm}, suggests that our universe encompasses at least one mesoscopic extra dimension with a diameter ranging from $0.1$ to $10 ~ \mu\text{m}$. Motivated by the smallness of dark energy $\Lambda$, this extra dimension—dubbed the Dark Dimension—leads to the emergence of an infinite tower of states whose mass is associated with dark energy.  Combining theoretical constraints with experimental results, one concludes that a tower of light states should be a Kaluza-Klein (KK) tower of mass $m_{\text{KK}}\sim \Lambda^{1/4}$, signaling decompactification of only one extra dimension~\cite{Montero:2022prj}. In the Dark Dimension (DD) scenario, the Standard Model of particle physics should live in a three-dimensional brane localized in the large extra dimension leaving only gravity and particles that are singlet under the standard model group propagating in the higher-dimensional bulk. It provides a bridge from quantum gravity to cosmology and particle physics; for a review see, for example,~\cite{Anchordoqui:2022svl} and references therein. 

In particular, micron-size extra dimensions offer an alternative framework that can explain the smallness of neutrino masses by postulating that the right-handed (R) neutrinos propagate in the dark dimension, in addition to the graviton \cite{Dienes:1998sb,Arkani-Hamed:1998wuz,Dvali:1999cn}. For an extra dimension that opens up at the micrometer region associated with the dark energy scale, the 5D Planck mass is of order $10^9$ GeV. In contrast, the conventional see-saw mechanism on the three-brane requires a 4D Majorana fermion with a mass around $10^{13}-10^{14}$ GeV, which is much heavier than the 5D Planck mass. However, if there are R-neutrinos propagating in the bulk, their coupling with the left-handed lepton doublet and the electroweak Higgs boson localized in the brane acquires a natural wave function suppression yielding naturally light Dirac neutrinos.

When the bulk masses of five-dimensional R-neutrinos vanish, recent analysis of neutrino oscillation data suggested that there exists an upper bound on the compactification radius around $0.4\,\mu\text{m}$ for Normal Hierarchy (NH) and $0.2\,\mu\text{m}$ for Inverted Hierarchy (IH)~\cite{Machado:2011jt,Forero:2022skg}.\footnote{More references on constraining extra dimensional models with bulk massless fermions by utilizing neutrino experiments can be found in~\cite{Barbieri:2000mg, Mohapatra:2000wn, Esmaili:2014esa, DiIura:2014csa, Berryman:2016szd, Basto-Gonzalez:2021aus, Khan:2022bcl, Roy:2023dyq, Siyeon:2024pte, Panda:2024ioo, Elacmaz:2025ihm}.} These upper bounds can be relaxed when the R-neutrinos are massive~\cite{Lukas:2000wn,Agashe:2000rw,Lukas:2000rg,Diego:2008zu,Carena:2017qhd,Anchordoqui:2023wkm,Eller:2025lsh,Antoniadis:2025rck,deGiorgi:2025xgp}. In particular, in the case with large positive bulk masses, where the lightest active neutrino mass is suppressed exponentially~\cite{Antoniadis:2025rck,deGiorgi:2025xgp}, there is no theoretical upper bound on the compactification radius. It can be as large as $30\,\mu\text{m}$, which is the model independent experimental upper bound from short distance measurements of Newton’s gravitational inverse-square law. In this work, we focus on the experimental signatures of massive right-handed neutrinos propagating along the extra direction of size as large as $10\,\mu\text{m}$ that may be observed in long-baseline neutrino oscillation experiments such as T2K and NOvA. 
Thus, we will consider only the parameter regions where the bulk masses of the R-neutrinos are larger than the compactification scale. 

The outline of this paper is as follows. In Section 2, we start by a short review of our model that provides the mechanism of generating neutrino masses by introducing three R-neutrinos, one for each lepton generation. The transition probabilities for neutrino oscillation in vacuum and in matter are discussed in subsections 2.2 and 2.3, respectively. Properties of the mass eigenvalues, mixing coefficients, and other theoretical ingredients are also provided. In Section 3, we present our analysis method in detail and show the results. We reserve Section 4 for our conclusions and outlook.

\section{Extra dimensional models of massive neutrinos}

\subsection{Overview of the theoretical model}

Our model lives in a 5D spacetime
that is the direct product of the 4D Minkowski spacetime labeled by coordinates $x^\mu$ and an extra direction that is an interval of length $\pi R$ for which we use $z\in[0,\pi R]$ as its coordinate.\footnote{The interval can be interpreted as the $\bbZ_2$ orbifold of a circle of radius $R$.} This spacetime can be pictured as the interval attached to every point of the 4D flat spacetime. We suppose that the end $z=0$ is attached to the 4D spacetime. The model contains three massive bulk fermions that are 5D Dirac spinors, and the Standard Model (SM) fields localized in the 4D spacetime. Therefore, the 4D spacetime is also called the SM (three-)brane. The bulk fermions are coupled with the left-handed leptons in the SM brane and the coupling sources masses of the SM neutrinos. 

The action of our model consists of three parts: the bulk action $S_\rb$, the bulk-brane interaction $S_{\rb\pd}$ and the SM action $S_{\trm{SM}}$. The bulk action is given by\footnote{We use the notation of~\cite{Antoniadis:2025rck}. The latin index $M$ refers to the five coordinates $0,1,\cdots,4$ with identification $x^4\equiv z$ while we use $\mu$ for the Minkowski directions $0,1,2,3$. 5D Dirac spinors have the same size as 4D Dirac ones. The 4D part of the 5D Dirac matrices $\Ga^\mu$ is identified with the 4D Dirac matrices, and $\Ga^4$ is chosen to be proportional to the 4D chirality matrix $\ga$. The left- and right-handed parts of $\Psi_i$ are defined by $\Psi_{\rL,i}=(1-\ga)\Psi_i/2$ and $\Psi_{\rR,i}=(1+\ga)\Psi_i/2$, respectively, coinciding with the chirality in 4D.}
\begin{align}
    S_\rb=\int\!d^4xdz\,\sum_{i=1}^3\,(i\bar\Psi_i\Ga^M\pd_M\Psi_i-c_i\bar\Psi_i\Psi_i),
\end{align}
where the fields $\Psi_i(x,z)$ ($i=1,2,3$) are the 5D Dirac spinors and the constants $c_i$ stand for their masses. The small Latin letters $i,j$ refer to the so-called intermediate basis, as explained shortly. The bulk-brane interaction is a Yukawa-type interaction of the right-handed part of bulk fermions $\Psi_{\rR,i}$ with the SM Higgs doublet and the left-handed lepton doublets:
\begin{align}
    S_{\rb\pd}=-M_*^{-\frac{1}{2}}\int\!d^4x\,\sum_{i=1}^3\,[y_i\ol{\ell_{\rL,i}}\tilde H\Psi_{\rR,i}(z=0)+\trm{c.c.}],
\end{align}
where $M_*$ is the 5D Planck mass and the constants $y_i$ are dimensionless bulk-brane coupling constants. $\tilde H$ is the conjugated Higgs doublet that acquires the vacuum expectation value $(v,0)$ with $v=174$\,GeV upon the electroweak symmetry breaking, and $\ell_{\rL i}$ stands for the left-handed lepton doublet in the intermediate basis. This basis is characterized as the basis in which the bulk-brane interaction is diagonal, distinct from the flavor basis labeled by small Greek letters $\al,\bt$ in which the weak interaction of the lepton doublets with the weak bosons has a diagonal form in the SM action $\ol{\nu_{\rL,\al}}\ga^\mu e_{\rL,\al}W^+_\mu$. The leptons in the two bases are related by~\cite{Machado:2011jt,Carena:2017qhd}
\begin{align}
    \ell_{\rL,\al}=\sum_iU_{\al i}\ell_{\rL,i}
\end{align}
with a $3\times3$ unitary matrix $U$, which plays a similar role to the Pontecorvo-Maki-Nakagawa-Sakata (PMNS) matrix.

To interpret the 5D model from the 4D perspective, we introduce mode functions on the interval, called KK mode functions, and expand the bulk fermions in them. We fix the mode functions by imposing boundary conditions at the two endpoints of the interval which require that the left-handed part of the bulk fermions should vanish at the endpoints:\footnote{Note that the left- and right-handed parts of the bulk fermions are correlated at the two endpoints and hence it is over-constraining to impose a boundary condition on both-handed parts.}
\begin{align}
    \Psi_{\rL,i}(z=0)=\Psi_{\rL,i}(z=\pi R)=0,
\end{align}
which removes the zero mode for the left-handed part $\Psi_{\rL,i}$ while keeping that for the right-handed part $\Psi_{\rR,i}$. Let us denote the left-handed mode functions by $f_{\rL,in}$ with positive integers $n=1,2,\cdots$, and the right-handed ones by $f_{\rR,in}$ with non-negative integers $n=0,1,2,\cdots$. We call the integer label $n$ the KK label. The bulk fermions are then expanded in the mode functions as $\Psi_{\rL/\rR,i}(x,z)=\sum_n\psi_{\rL/\rR,in}(x)f_{\rL/\rR,in}(z)$, and the coefficients $\psi_{\rL/\rR,in}$ behave as the 4D fermions. Plugging the expansions into the actions $S_\rb+S_{\rb\pd}$ and using the orthonormality of the mode functions, we obtain
\begin{align}\label{S4}
    \begin{split}
    S_4=\int d^4x&\sum_i\,\Big[i\ol{\nu_{\rL,i}}\ga^\mu\pd_\mu\nu_{\rL,i}+\msum_{n\geq1}i\ol{\psi_{\rL,in}}\ga^\mu\pd_\mu\psi_{\rL,in}+\msum_{n\geq0}i\ol{\psi_{\rR,in}}\ga^\mu\pd_\mu\psi_{\rR,in} \\
    &-\msum_{n\geq1}\lm_{in}(\ol{\psi_{\rR,in}}\psi_{\rL,in}+\ol{\psi_{\rL,in}}\psi_{\rR,in})-\msum_{n\geq0}(Y_{in}\ol{\nu_{\rL,i}}\psi_{\rR,in}+Y_{in}^*\ol{\psi_{\rR,in}}\nu_{\rL,i})\Big],
    \end{split}
\end{align}
where we included the kinetic term of the SM left-handed neutrinos $\nu_{\rL,i}$ from the SM action, and introduced
\begin{align}\label{lmY}
    \lm_{in}=\sqrt{c_i^2+\frac{n^2}{R^2}}, \quad
    Y_{i0}=\mu_i\sqrt{\frac{2\pi c_iR}{e^{2\pi c_iR}-1}}, \quad
    Y_{in}=\mu_i\sqrt{\frac{2n^2}{n^2+c_i^2R^2}} \quad (n\geq1),
\end{align}
together with the parameters $\mu_i$ of mass dimension one, defined as rescaled bulk-brane coupling constants $y_i$:
\begin{align}
    \mu_i=M_*^{-\frac{1}{2}}\frac{vy_i}{\sqrt{\pi R}}.
\end{align}
Our model is therefore characterized by the unitary matrix $U$ introduced above and the seven parameters
\begin{align}
(R,c_1,c_2,c_3,\mu_1,\mu_2,\mu_3).
\end{align}

\subsection{Oscillation in vacuum}
The mass terms in the 4D action \eqref{S4} are of Dirac type. In particular, SM neutrinos $\nu_{\rL,i}$ are coupled with the right-handed zero modes $\psi_{\rR,i0}$ as well as higher modes $\psi_{\rR,in}$ $(n\geq1)$, which makes the mass terms not diagonal. We therefore need to diagonalize the mass term in order to read off the physical masses.
The mass term can be rewritten in a matrix form:
\begin{align}
    \int d^4x \msum_i\msum_{m,n\geq0}\ol{\psi_{\rL,im}}(\cM_i)_{mn}\psi_{\rL,in},
\end{align}
where the matrix $\cM_i$ indexed by KK labels $mn$ is defined by
\begin{align}
    &(\cM_i)_{0n}=Y_{in} \quad (n\geq0), \\ 
    &(\cM_i)_{nn}=\lm_{in} \quad (n\geq1),
\end{align}
with the other matrix elements being zero.
The Schr\"odinger equation for 4D fermions $\psi_{in}$ of energy $E$ in the intermediate basis is given by
\begin{align}\label{Scheq-vac}
    i\frac{dc_{in}(t)}{dt}=\sum_{m\geq0}\frac{(\cM_i\cM_i^\dg)_{nm}}{2E}c_{im}(t).
\end{align}
It is solved by diagonalizing the mass square matrix $\cM_i\cM_i^\dg$. Its diagonalization reads
\begin{align}
    (\cM_i\cM_i^\dg)_{nm}=\msum_{\ell\geq0}\cL^i_{n(\ell)}m^2_{i(\ell)}\cL^{i*}_{m(\ell)}, \quad
    \msum_{n\geq0}\cL^{i*}_{n(\ell)}\cL^i_{n(\ell')}=\de_{\ell\ell'},
\end{align}
where $m^2_{i(\ell)}$ $(\ell=0,1,2\cdots)$ are the eigenvalues of $\cM_i\cM_i^\dg$ in the increasing order, and the vector $(\cL^i_{0(\ell)},\cL^i_{1(\ell)},\cL^i_{2(\ell)},\cdots)$ is the unit eigenvector for eigenvalue $m^2_{i(\ell)}$. A general solution to \eqref{Scheq-vac} is given by
\begin{align}
    c_{in}(t)=\sum_{\ell,m\geq0}\cL^i_{n(\ell)}\exp\left(-i\frac{t}{2E}m^2_{i(\ell)}\right)\cL^{i*}_{m(\ell)}c_{im}(0).
\end{align}
The wavefunction for the SM neutrino in the intermediate basis is $c_{i0}$ and related to that in the flavor basis by $c_{\al 0}=\sum_iU_{\al i}^*c_{i0}$. 
The probability that a neutrino of flavor $\al$ at time 0 is found to be of flavor $\bt$ at time $t$ is then given by $|c_{\bt0}(t)|^2$ upon the initial condition $c_{\ga n}(0)=\de_{\ga\al}\de_{n0}$, which reads explicitly
\begin{gather}
    P_{\al\to\bt}(E,t)=|A_{\al\bt}(E,t)|^2, \\
    A_{\al\bt}(E,t):=c_{\bt0}(t)=\sum_{i}\sum_{\ell\geq0}U^*_{\al i}\cL^{i*}_{0(\ell)}\exp\left(-i\frac{t}{2E}m^2_{i(\ell)}\right)U_{\bt i}\cL^i_{0(\ell)}.
\end{gather}
One can see from this that $\sum_iU_{\al i}\cL^i_{0(\ell)}$ plays a role of the mixing between the flavor and mass bases.
Properties of the characteristic equation of $\cM_i\cM_i^\dg$ will be presented later.

\subsection{Oscillation in matter}
Let us proceed to the oscillation probability with matter effect. Neutrinos make weak interaction with electrons and nuclei in matter during propagation and feel an effective weak potential. 
This effect can be expressed as a change in the kinetic term of the SM neutrino:
\begin{align}
    \msum_i\ol{\nu_{\rL,i}}\ga^\mu i\pd_\mu\nu_{\rL,i} ~~\to~~\msum_{ij}\ol{\nu_{\rL,i}}\ga^\mu(i\pd_\mu\de_{ij}-V_{\mu,ij})\nu_{\rL,j},
\end{align}
where $V_{\mu,ij}$ is the effective weak potential. In the setup of our interest, it is reasonable to assume that $V_0$ is the only non-vanishing component, reading~\cite{Machado:2011jt}
\begin{align}\label{Vij}
    V_{0,ij}=\sum_\al U_{\al i}^*G_F\left(\sqrt{2}n_e\de_{\al e}-\frac{1}{\sqrt 2}n_n\right)U_{\al j},
\end{align}
where $G_F$ is the Fermi constant and $n_e$$(n_n)$ is the number density of the electron (neutron) in matter. This changes the Schr\"odinger equation in vacuum \eqref{Scheq-vac} to
%\footnote{See for example~\cite{Akhmedov:2020vua} for a derivation.}
\begin{align}
    i\frac{dc_{in}(t)}{dt}=\sum_j\sum_{m\geq0}\left[\frac{(\cM_i\cM_i^\dg)_{nm}}{2E}\de_{ij}+\de_{n0}\de_{m0}V_{0,ij}\right]c_{jm}(t).
\end{align}
It is solved by diagonalizing a matrix $\sfH$ defined by
\begin{align}
    \sfH_{in,jm}=(\cM_i\cM_i^\dg)_{nm}\de_{ij}+2E\de_{n0}\de_{m0}V_{ij}.
\end{align}
In contrast to the vacuum case, the matrix is not diagonal in $ij$. For concrete computation, it is convenient to introduce a cutoff $N$ to KK labels so that indices $n,m$ refer to $0,1,\cdots,N$ and the matrix $\sfH$ is considered to have size $(3N+3)\times(3N+3)$ under the identification of index pair $in$ with a single index $i+3n$. The diagonalization of $\sfH$ then reads
\begin{align}\label{Hdef}
    \sfH_{in,jm}=\msum_{h=1}^{3N+3}\sfL_{in(h)}m^2_{(h)}\sfL^\dg_{(h)jm},
\end{align}
where $m^2_{(h)}$ $(1\leq h\leq3N+3)$ are the eigenvalues of $\sfH$ in the increasing order, and $\sfL_{in(h)}$ is the $(i+3n)$-component of the normalized eigenvector of size $3N+3$ for eigenvalue $m_{(h)}^2$.  
Properties of the eigenvalues and eigenvectors will be given in the next subsection. Once they are obtained, the oscillation probability $P_{\al\to\bt}(E,t)$ is given by\footnote{The oscillation probability for antineutrinos $P_{\bar\al\to\bar\bt}$ is obtained by replacing $V_{0,ij}$ by $-V_{0,ij}$ and $U_{\al i}$ by $U_{\al i}^*$.}
\begin{gather}
    P_{\al\to\bt}(E,t)=|A_{\al\bt}(E,t)|^2, \\
    A_{\al\bt}(E,t)=\sum_{ij}\sum_{h=1}^{3N+3}U^*_{\al i}\sfL^*_{i0(h)}\exp\left(-i\frac{t}{2E}m^2_{(h)}\right)U_{\bt j}\sfL_{j0(h)}.
\end{gather}
In Section~\ref{sec:simulations}, we will compute the oscillation probability by obtaining the mass eigenvalues in matter by numerical diagonalization of $\sfH$ \eqref{Hdef}. Proper choice of the matrix size (mode cutoff) will be discussed there as well as in Section~\ref{subsec:properties}.

\subsection{Mass eigenvalues and mixing coefficients}

The mass eigenvalues are obtained by solving the characteristic equation
\begin{align}\label{chareq1}
    \det(R^2\sfH-x\sfI)=0,
\end{align}
where $\sfI$ is the identity matrix.\footnote{Here both $\sfH$ and $\sfI$ are regarded as matrices of size $(3N+3)\times(3N+3)$ under the identification of $in$ with $i+3n$ as explained in the last subsection.} We multiplied $\sfH$ by $R^2$ to make it dimensionless and introduced a dimensionless variable $x$ so that eigenvalue $m^2_{(h)}$ of $\sfH$ corresponds to a root $x_{(h)}=R^2m_{(h)}^2$ of \eqref{chareq1}.
This equation can be rewritten by using the standard Gaussian elimination in terms of the determinant of a 3$\times$3 matrix $T(x)$:\footnote{This is a generalization of the derivation in~\cite{Machado:2011jt} where the case with massless bulk fermions was considered.}
\begin{gather}
    0=\det T(x)\cdot\prod_{h=1}^{3N+3}(R^2\lm_{in}^2-x), \label{chareqT} \\
    T(x)=\begin{pmatrix}
        t_{1}(x)+v_{11} & v_{12} & v_{13} \\
        v_{21} & t_{2}(x)+v_{22} & v_{23} \\
        v_{31} & v_{32} & t_{3}(x)+v_{33}
    \end{pmatrix},
\end{gather}
where $t_i(x)$ and $v_{ij}$ are defined by
\begin{align}\label{tNi}
    t_i(x)=\sum_{n=0}^N|RY_{in}|^2+\sum_{n=1}^N\frac{|RY_{in}|^2(R\lm_{in})^2}{x-(R\lm_{in})^2}-x, \quad v_{ij}=2R^2EV_{ij}.
\end{align}
Since $x=R^2\lm_{in}^2$ cannot solve the equation\footnote{It is because $t_i$ then becomes $\sim(x-R^2\lm_{in}^2)^{-1}$ and this cancels $R^2\lm_{in}^2-x$ from the product in \eqref{chareqT}.}, it is sufficient to solve $\det T(x)=0$ in order to find all eigenvalues. In the $N\to\infty$ limit, the infinite series in \eqref{tNi} can be resummed into
\begin{align}\label{coteq}
    t_i(x)=\pi\bar\mu_i^2\sqrt{x-\bar c_i^2}\cot(\pi\sqrt{x-\bar c_i^2})-x-\pi\bar\mu_i^2\bar c_i,
\end{align}
where we introduced dimensionless parameters $\bar c_i,\bar\mu_i$ by rescaling $c_i$ and $\mu_i$ with the compactification radius $R$ as in~\cite{Antoniadis:2025rck}:
\begin{align}
    \bar c_i=c_iR, \quad \bar\mu_i=\mu_iR.
\end{align}
An important property is that $t_i(x)$ is well-defined even for $x<\bar c_i^2$ and $\cot$ is replaced with its hyperbolic counterpart $\coth$~\cite{Anchordoqui:2023wkm,Antoniadis:2025rck}:
\begin{align}\label{cotheq}
    t_i(x)=\pi\bar\mu_i^2\sqrt{\bar c_i^2-x}\coth(\pi\sqrt{\bar c_i^2-x})-x-\pi\bar\mu_i^2\bar c_i \quad (x<\bar c_i^2).
\end{align}
This coth branch is a distinctive feature compared with the case with massless bulk fermions.

The eigenvalue equation and the orthonormality of the eigenvectors are given by
\begin{align}
    &\msum_{jm}R^2\sfH_{in,jm}\sfL_{jm(h)}=x_{(h)}\sfL_{in(h)}, \label{eveq} \\
    &\msum_{in}\sfL_{in(h)}\sfL_{in(h')}^*=\de_{hh'}. \label{on}
\end{align}
One can show that the eigenvalue equation \eqref{eveq} is equivalent to
\begin{align}
    &\msum_jT(x_{(h)})_{ij}\sfL_{j0(h)}=0, \label{Li0eq} \\
    &\sfL_{in(h)}=\frac{RY_{in}\cdot R\lm_{in}}{x_{(h)}-(R\lm_{in})^2}\sfL_{i0(h)} \quad (n\geq1). \label{Lineq}
\end{align}
One can see that once three $\sfL_{i0(h)}$ are given, the others $\sfL_{in(h)}$ are determined by \eqref{Lineq}. Since $\det T(x_{(h)})=0$, the three linear equations \eqref{Li0eq} are not linearly independent and their solutions contain one free parameter, which can be fixed by the normalization condition \eqref{on}. Therefore, a procedure to find $\sfL_{i0(h)}$ is first to solve
\begin{align}
    \msum_iT_{2i}(x_{(h)})w_{i}=0, \quad \msum_iT_{3i}(x_{(h)})w_{i}=0
\end{align}
for $w_2,w_3$ with $w_1=1$, and then to rescale $w_{1},w_{2},w_{3}$ to make them satisfy the normalization condition. The result is:
\begin{align}
    &\sfL_{10(h)}=\left(\msum_i|w_{i}|^2\ell_i(x_{(h)})\right)^{-\frac{1}{2}}, \quad
    \sfL_{20(h)}=\sfL_{10(h)}w_{2}, \quad
    \sfL_{30(h)}=\sfL_{10(h)}w_{3}, \\
    &w_{1}=1, \quad 
    w_{2}=\frac{T_{(h)23}T_{(h)31}-T_{(h)21}T_{(h)33}}{T_{(h)22}T_{(h)33}-T_{(h)23}T_{(h)32}}, \quad
    w_{3}=\frac{T_{(h)21}T_{(h)32}-T_{(h)22}T_{(h)31}}{T_{(h)22}T_{(h)33}-T_{(h)23}T_{(h)32}},
\end{align}
where $T_{(h)ij}=T(x_{(h)})_{ij}$ and $\ell_i(x)$ is defined by
\begin{align}
    \ell_i(x)=-\frac{dt_i(x)}{dx}.
\end{align}
The results about the eigenvectors obtained so far hold true whether $N$ is finite or infinite. 

In the vacuum case where $v_{ij}=0$, the mass eigenvalues $m_{i(\ell)}^2$ and the mixing coefficients $\cL^i_{0(\ell)}$ are given by~\cite{Anchordoqui:2023wkm,Antoniadis:2025rck} 
\begin{align}\label{masseqvac}
    t_i(R^2m_{i(\ell)}^2)=0, \quad \cL^i_{0(\ell)}=\ell_i(R^2m_{i(\ell)}^2)^{-\frac{1}{2}}.
\end{align}
Exactly speaking, setting $v_{ij}=0$ in the characteristic equation just implies a weaker condition $\prod_jt_j(R^2m_{i(\ell)}^2)=0$.
But we can derive $t_i(R^2m_{i(\ell)}^2)=0$ by applying the Gaussian elimination to $\cM_i\cM_i^\dg$ as we did to $\sfH$.

\subsection{Various properties}
\label{subsec:properties}
\paragraph{In vacuum.}
We review properties of the mass eigenvalues in vacuum based on~\cite{Anchordoqui:2023wkm,Antoniadis:2025rck}. For each $i$, the eigenvalues $x_{i(n)}=R^2m_{i(n)}^2$ $(n\geq0)$ are the positive roots of $t_i(x)=0$ as stated in \eqref{masseqvac}. We define label $(n)$ in $x_{i(n)}$ based on the region to which each $x_{i(n)}$ belongs. Its detail depends on the sign of $t_i(|\bar c_i|)=\bar\mu_i^2-\bar c_i^2-\pi\bar\mu_i^2\bar c_i$ as follows:
\begin{alignat}{2}
    t_i(|\bar c_i|)\leq0&:~~\quad
    &0\leq x_{i(0)}\leq|\bar c_i|^2, \quad &|\bar c_i|^2+n^2< x_{i(n)}<|\bar c_i|^2+(n+1)^2 \quad (n\geq1), \\
    t_i(|\bar c_i|)>0&:\quad  &{}
    &|\bar c_i|^2+n^2< x_{i(n)}<|\bar c_i|^2+(n+1)^2 \quad (n\geq0).
\end{alignat}
Namely, the sign of $t_i(|\bar c_i|)$ determines whether the zero mode $x_{i(0)}$ is smaller or larger than $\bar c_i^2$. If it is smaller than $\bar c_i^2$ it solves the coth equation \eqref{cotheq}, otherwise it solves the cot equation \eqref{coteq}.

The zero mode masses are identified with the masses $m_i$ of the SM neutrinos: $m_i=m_{i(0)}$.
Combining this with the mass square differences $\De m_{i1}^2=m_i^2-m_1^2$, we can fix two parameters out of three $\bar\mu_i$~\cite{Forero:2022skg,Antoniadis:2025rck}.
Let us suppose that the neutrino masses are in NH. We first obtain the smallest neutrino mass $m_1$ from the smallest root of $t_1(x)=0$. Once the mass square differences $\De m_{21}^2,\De m_{31}^2$ are available, we can determine the other masses by $m_i^2=m_1^2+\De m_{i1}^2$. We then identify them with the smallest roots of the mass equations $t_2(x)=t_3(x)=0$ as we did for $m_1$; namely $m_2=m_{2(0)}$ and $m_3=m_{3(0)}$. Requiring $t_2(R^2m_2^2)=t_3(R^2m_3^2)=0$ fixes $\bar\mu_2^2,\bar\mu_3^2$ as
\begin{align}\label{mu23}
\begin{split}
    \bar\mu_i^2 &= \frac{R^2m_i^2}{\pi\sqrt{\bar c_i^2-R^2m_i^2}\coth(\pi\sqrt{\bar c_i^2-R^2m_i^2})-\pi\bar c_i} \qquad (R^2m_i^2\leq\bar c_i^2), \\
    \bar\mu_i^2 &= \frac{R^2m_i^2}{\pi\sqrt{R^2m_i^2-\bar c_i^2}\cot(\pi\sqrt{R^2m_i^2-\bar c_i^2})-\pi\bar c_i} \qquad (R^2m_i^2>\bar c_i^2).
\end{split}
\end{align}
Since $\bar\mu_2$ and $\bar\mu_3$ are real numbers, we need to require the positivity $\bar\mu_2^2,\bar\mu_3^2>0$. On top of this, we have the constraints that $(Rm_2)^2,(Rm_3)^2$ are confined to the regions where $x_{2(0)},x_{3(0)}$ are located. We therefore find the following consistency conditions: 
\begin{align}
    R^2m_i^2\leq\bar c_i^2: &\quad \sqrt{\bar c_i^2-R^2m_i^2}\coth(\pi\sqrt{\bar c_i^2-R^2m_i^2})-\bar c_i>0, \\
    % \begin{split}
    R^2m_i^2>\bar c_i^2: &\quad \sqrt{R^2m_i^2-\bar c_i^2}\cot(\pi\sqrt{R^2m_i^2-\bar c_i^2})-\bar c_i>0 ~~\text{ and }~~ R^2m_i^2<|\bar c_i|^2+1.
    % \end{split}
\end{align}
One can see that negative $\bar c_2,\bar c_3$ are convenient for the positivity though this is not a sufficient condition.\footnote{It is satisfied when the bulk masses are ``large'' in the sense of the condition \eqref{largec} discussed later.} This was also inferred in~\cite{Antoniadis:2025rck} by requiring the above constraints together with the compactification radius of order 1--$10\,\mu\mrm{m}$ proposed in the DD scenario and the cosmological upper bounds on the neutrino masses~\cite{Planck:2018nkj,Planck:2018vyg}. Following them, this article will adopt this choice $c_2,c_3<0$. 

In summary, the identification of the active neutrino masses with the smallest roots of the mass equations upon the use of the mass square differences reduce the parameters by two to
\begin{align}
    R,~~\bar c_1,~~\bar c_2,~~\bar c_3,~~\bar\mu_1.
\end{align}
The construction goes in a parallel manner in the IH case by obtaining $m_3$ instead of $m_1$ as the smallest root of $t_3(x)=0$ parameterized by $\bar c_3$ and $\bar\mu_3$, so that $\bar\mu_1,\bar\mu_2$ are fixed through the mass square differences in IH and thus the free parameters are $R,\bar c_1,\bar c_2,\bar c_3,\bar\mu_3$.

\begin{figure}[htb!]
    \centering
    \includegraphics[width=1.0\linewidth]{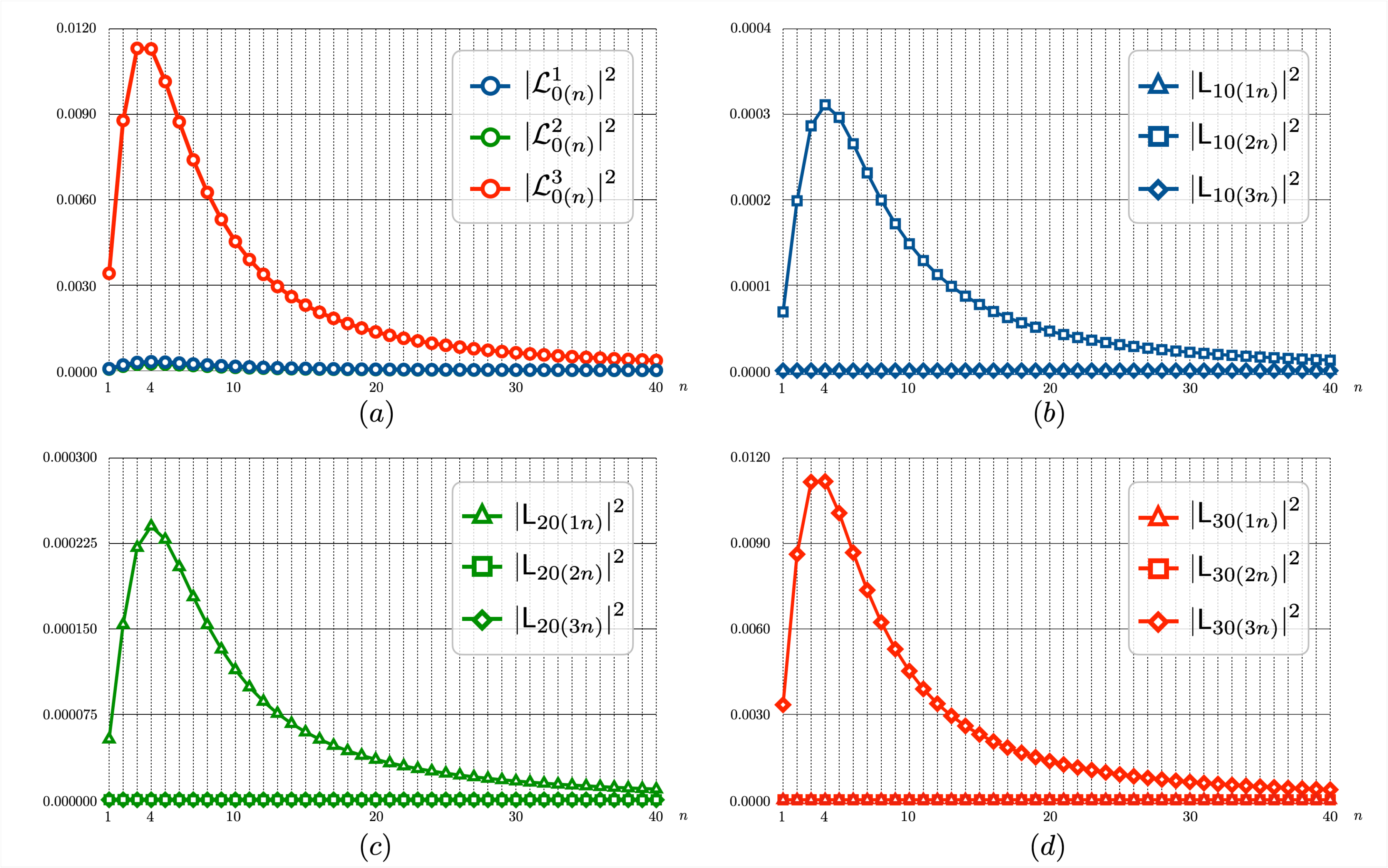}
    \caption{Panel $(a)$ plots the values of the squared mixing coefficients in vacuum $|\mathcal{L}^i_{0(n)}|^2$ for $n = 1, \dots ,40$. Panels $(b)$, $(c)$, $(d)$ plots the values of the squared mixing coefficients $|\mathsf{L}_{10(kn)}|^2$, $|\mathsf{L}_{20(kn)}|^2$, $|\mathsf{L}_{30(kn)}|^2$, respectively. Parameters for the plots are $\bar c_1=-\bar c_2 = -\bar c_3 = 4$ and $\bar \mu_1 = 0.1$ with $\bar\mu_2,\bar\mu_3$ determined by \eqref{mu23} in NH.}
    \label{Mixing_coff}
\end{figure}

\paragraph{In matter.}
Let us consider the case in matter. 
In this article we will focus on the case where the three bulk masses have the same absolute value $\bar c:=|\bar c_1|=|\bar c_2|=|\bar c_3|$ as in~\cite{Antoniadis:2025rck}. In this case, since the cot/coth functions in $\det T(x)$ contain the common factor $\pi\sqrt{|\bar c^2-x|}$, the roots are separated by $\bar c^2+n^2$ where cot functions and thus $\det T$ diverge, and each region made by the separation contains three roots. Therefore, the roots of $\det T(x)=0$ can be labeled as $x_{(kn)}$ with $1\leq k\leq 3$ and $0\leq n\leq N$ in the following way:
\begin{align}
    &x_{(10)}<x_{(20)}<x_{(30)}\leq|\bar c|^2+1, \\
    |\bar c|^2+n^2<~&x_{(1n)}<x_{(2n)}<x_{(3n)}<|\bar c|^2+(n+1)^2 \quad (n\geq1),
\end{align}
where $k$ refers to three roots in each region specified by $n$.
Geometrically, the continuous curve in vacuum $\prod_it_i(x)$ in each region with three zeros is continuously deformed by $v_{ij}$ into another curve in the same region keeping the same number of zeros. 

\paragraph{Mixing.} The oscillation probability formula in our model involves an infinite sum over KK modes. Therefore, its numerical computation needs a cutoff. Proper choice of the cutoff can be seen from the behavior of the mixing coefficients. In vacuum, the mixing coefficient $\cL^i_{0(n)}$ for each $i$ has the maximum at zero mode. However, it was demonstrated~\cite{Carena:2017qhd,Anchordoqui:2023wkm,Antoniadis:2025rck,deGiorgi:2025xgp} that if the zero mode satisfies the coth equation ($m_{i(0)}<|c_i|$), the mixing coefficient has another peak that is much smaller than the zero-mode peak. This is demonstrated in Figure~\ref{Mixing_coff}(a) for $\bar c=4$, where the three coefficients indeed have peaks at $n=4$. This feature is kept in matter in the experimental setups of this paper (T2K, NOvA). Figure~\ref{Mixing_coff}(b), (c), (d) plot the mixing coefficients $\mathsf{L}_{10(kn)},\mathsf{L}_{20(kn)},\mathsf{L}_{30(kn)}$, respectively, for $\bar c=4$, where one can see that each has a peak at $n=4$.

The presence of the second peak suggests that the cutoff on the infinite sum in the oscillation probability should be chosen to be larger than the position of the second peak around $n\sim|\bar c|$ in order to capture the contribution from the second peak. The probability decays as $\sim1/\cN$ as long as a cutoff $\cN$ on the KK mode sum is chosen this way. The behavior of the probability under various cutoffs for $\bar c=4$ are demonstrated in Figure~\ref{cutoff}, which implies that cutoffs just near the peak of the mixing $n\sim4$ do not approximate the probability with the infinite sum sufficiently. It is because the value of the second peak is much smaller than the global peak at zero mode and the other values around the second peak are as small as it due to the slow decay as $\cL^i_{0(n)}\sim1/n$. 

\paragraph{Large bulk mass.}
In~\cite{Antoniadis:2025rck}, the following parameter region with large bulk masses:
\begin{align}\label{largec}
    |\bar c_i|\gg\max\{1,\bar\mu_i,\pi\bar\mu_i^2\}
\end{align}
was investigated in the vacuum case. In this region, we can write down the physical masses and the mixing coefficients analytically as
\begin{align}
    &m_{i(0)}\simeq\begin{cases} \mu_i\sqrt{2\pi|\bar c_i|}e^{-\pi\bar c_i} \qquad \text{when }~c_i>0 \\
    \mu_i\sqrt{2\pi|\bar c_i|} \qquad\qquad~ \text{when }~c_i<0
    \end{cases}, \label{mi0-3+1} \\
    &m_{i(n)}\simeq\sqrt{c_i^2+\frac{n^2}{R^2}}\left[1+\frac{n^2\bar\mu_i^2}{(\bar c_i^2+n^2)^2}+\cdots\right] \quad (n\geq1), \\
    &\cL^i_{0(0)}\simeq1-\frac{\pi\bar\mu_i^2}{4|\bar c_i|}, \qquad
    \cL^i_{0(n)}\simeq\frac{\sqrt{2}\bar\mu_in}{\bar c_i^2+n^2} \quad (n\geq1).
\end{align}
The zero mode masses are exponentially suppressed when $c_i>0$ and higher mode masses are pushed to the vicinity of the KK masses $\lm_{in}$ before mass diagonalization at which the mass equations diverge. The mixing coefficient $\cL^i_{0(n)}$ has the second peak at $n\sim|\bar c_i|$. Since the mass gap between higher KK modes is $\sim1/|\bar c_i|$, the spectrum of higher KK modes looks continuous for large $|\bar c_i|$. Since the mixing coefficients are peaked around $n\sim|\bar c_i|$, the mass spectrum will approach that consisting of three zero modes and three continuous spectra with minimum masses $|c_i|$ ($i=1,2,3$). %that make the biggest contribution in the mixing coefficients. 
In the particular case of equal bulk masses $c:=|c_1|=|c_2|=|c_3|$, it was proven~\cite{Antoniadis:2025rck} by explicit computation\footnote{Technically, an infinite sum over KK modes can be replaced in the large bulk mass region by an integral over the ratio of KK modes to the bulk mass.} that the differential beta-decay spectrum of molecular tritium $\mrm{T}_2$ in the parameter region \eqref{largec} is well approximated by that in the so-called $3+1$ model with a sterile neutrino of mass $c$. This feature allowed the test of the parameter region \eqref{largec} by making good use of the results for sterile neutrinos in KATRIN~\cite{KATRIN:2025lph}. Note also that in the limit $\bar c:=cR\to\infty$ with finite $\bar\mu_i$, the continuum of the KK modes is infinitely gapped from the zero modes and hence the model approaches the 4D standard oscillation scenario.

In this paper, we will consider a parameter region where the condition \eqref{largec} is not necessarily valid, covering $\bar c\in[2.5,10]$, with the compactification radius $R=10\,\mu$m as a representative scale proposed in the DD scenario. Smaller values of $\bar c$ have been considered in~\cite{Carena:2017qhd,Eller:2025lsh,deGiorgi:2025xgp}, where compactification radii of order 10\,$\mu\text{m}$ were shown to be disfavored by using neutrino oscillation data, which is consistent with upper bounds on $R$ for such small values of $\bar c$ theoretically obtained in~\cite{Antoniadis:2025rck}. On the other hand, for larger values of $\bar c$ including order 1000 considered in~\cite{Antoniadis:2025rck}, the numerical diagonalization for mass eigenvalues that we adopt in this paper will encounter computational issues, and hence it is more effective to solve the mass equations in the infinite size limit \eqref{coteq} and \eqref{cotheq} numerically, which was employed in~\cite{Antoniadis:2025rck} for numerical computations of the beta-decay spectrum. Testing such a large $\bar c$ region including the region \eqref{largec} by using neutrino oscillation data, as well as evaluating the continuous (integral) limit of the oscillation probability formula, is ongoing work and will be reported in a forthcoming paper.

\begin{table}
    \centering
    \begin{tabular}{|c|c|}
    \hline
         Parameter & Value  \\ \hline
         $\sin^2\theta_{12}$ (NH) & {0.307 $\pm$ 0.012} \\ \hline
         $\sin^2\theta_{12}$ (IH) & {0.308 $\pm$ 0.012} \\ \hline
         $\sin^2\theta_{13}$ (NH) & {0.02195 $\pm$ 0.00056} \\ \hline
         $\sin^2\theta_{13}$ (IH) & {0.02224 $\pm$ 0.00057} \\ \hline
         $\sin^2\theta_{23}$ (NH) & $0.561\pm0.014$ \\ \hline
         $\sin^2\theta_{23}$ (IH) & $0.562\pm0.014$ \\ \hline
         $\delta_{\rm CP}$ $( \degree)$ (NH) & $177\pm20$ \\ \hline
         $\delta_{\rm CP}$ $( \degree)$ (IH) & $285\pm27$ \\ \hline
         $\Delta m_{21}^2$ $(10^{-5}~\rm{eV}^2)$ & $7.49 \pm 0.19$ \\ \hline
         $\Delta m_{31}^2$ $(10^{-3}~\rm{eV}^2)$ (NH) & $2.534\pm0.025$  \\ \hline
         $\Delta m_{32}^2$ $(10^{-3}~\rm{eV}^2)$ (IH) & $-2.510\pm0.025$ \\ \hline
    \end{tabular}
    \caption{ The global best fit values and 1$\sigma$ uncertainty of the standard oscillation parameters used in our study. The central values and 1$\sigma$ range of these parameters are taken from the latest global analysis NuFIT 6.0~\cite{Esteban:2024eli}.
    }
    \label{stdpara}
\end{table}

\section{Simulations}
\label{sec:simulations}
 We will conduct the analysis of T2K and NOvA data based on the likelihood test with $\chi^2$ in order to probe the difference between the standard and Dark Dimension (DD) scenarios. This work is done with the General Long Baseline Experiment Simulator (GLoBES) \cite{Huber:2004ka,Huber:2006xx}, which has been modified for our neutrino mass model based on the DD scenario. Detailed computational procedures and their implementations can be accessed through the public GitHub repository\cite{repolink}, where we use the experimental setup files updated from \cite{Lin:2023xyk}. We first introduce our parameter settings, the details of T2K and NOvA experiments and the $\chi^2$ function used in the analysis, and finally present our simulation results.
\subsection{Parameter Settings}
\label{settings}
To start with the simulation, we choose the central values and 1$\sigma$ uncertainties of the standard oscillation parameters from the latest global analysis NuFIT 6.0~\cite{Esteban:2024eli}. The standard oscillation parameters used in our study are shown in Table \ref{stdpara}. Note that in the simulations of T2K and NOvA experiments, we matched the best-fit values of $\theta_{23}$, $\delta_{\text{CP}}$, and $\De m_{32}^2$ to reproduce their published experimental results. We then examine the Dark Dimensional neutrino model based on the simulated neutrino oscillation spectra. The results obtained this way should be able to reflect the sensitivity of new physics induced by the DD neutrino model in the simulated data.

\begin{figure}[t]
    \centering
    \includegraphics[width=0.45\linewidth]{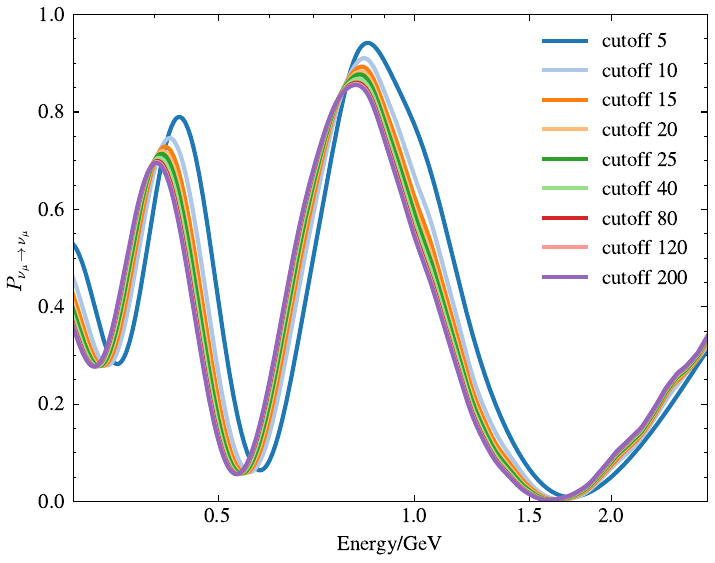}
\includegraphics[width=0.444\linewidth]{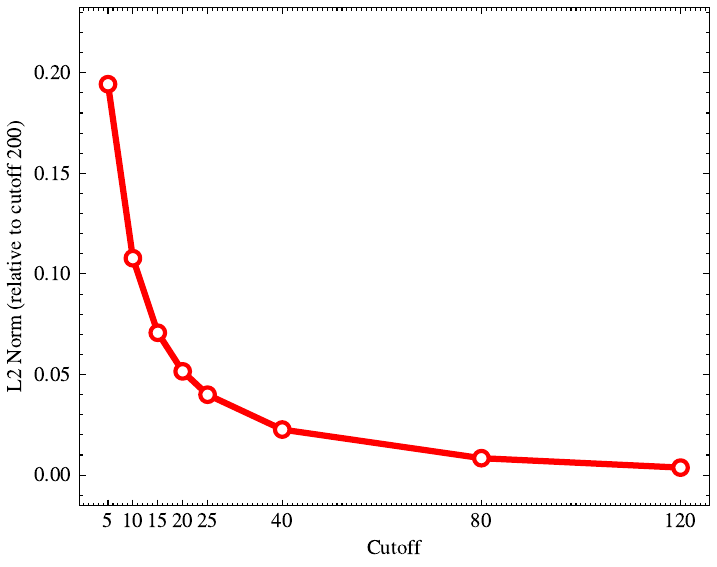}
    \caption{Left panel: Difference on $P_{\nu_\mu\to\nu_\mu}$ with different KK modes cutoff values when \{$R$, $|\bar{c}_i|$, $\bar{\mu}_1$\} are set as \{10~$\mu$m, 4, 0.1\} and the baseline of NOvA is used. The low-pass filter is used to smooth the curves, which is discussed in Section \ref{prob}. Right panel: the $L^2$ norm between $P_{\nu_\mu\to\nu_\mu}$ with cutoff $\in[5,120]$ and $P_{\nu_\mu\to\nu_\mu}$ with a fixed cutoff of 200. }
    \label{cutoff}
\end{figure}

Regarding the parameters of our model, we will investigate the case where $|\bar{c}_1|=|\bar{c}_2|=|\bar{c}_3|$ with $\bar{c}_1>0,~\bar{c}_2,\bar{c}_3<0$ for NH or $\bar{c}_3>0,~\bar{c}_1,\bar{c}_2<0$ for IH.\footnote{The case with three negative bulk masses is also possible. In particular, when they have the equal absolute value $c:=c_1=c_2=c_3<0$, which have been analyzed in~\cite{Eller:2025lsh,deGiorgi:2025xgp}, the lightest active neutrino mass is not suppressed exponentially but behaves as $\sim\mu_i\sqrt{|\bar c|}$ for large $|\bar c|$, as shown in \eqref{mi0-3+1}. Therefore, as $|\bar c|$ gets larger, the neutrino masses are more likely to reach their upper bound and thus the allowed parameter regions will be more restricted. Careful investigation of such cases with large $|\bar c|$ will be covered in our forthcoming work.} For NH, we set $\bar{\mu}_1$ as a free parameter. $\bar\mu_2$ and $\bar\mu_3$ are determined using \eqref{mu23}. Instead, we set $\bar\mu_3$ as a free parameter for IH and \eqref{mu23} is used to determine $\bar\mu_1$ and $\bar\mu_2$. 

Before proceeding, we make a remark about the cutoff dependence of the infinite KK mode sum in the neutrino oscillation probability computed in our simulation. A small cutoff will lead to incorrect results as pointed out in Section~\ref{subsec:properties}, while a too large cutoff will consume a significant amount of computing resources. Figure~\ref{cutoff} shows $P_{\nu_\mu\to\nu_\mu}$ with various cutoffs, where we choose NH and set \{$R$, $|\bar{c}_i|$, $\bar{\mu}_1$\} as \{10\,$\mu$m, 4, 0.1\}. One can also see that the $L^2$ norms\footnote{The $L^2$ norm between two one-variable functions $f,g$ is defined by $||f-g||=\left(\int dx|f(x)-g(x)|^2\right)^{1/2}$.} between $P_{\nu_\mu\to\nu_\mu}$ with a reference cutoff 200 and $P_{\nu_\mu\to\nu_\mu}$ with other cutoffs display a damping of $n^{-1}$, which is consistent with an estimate of the sum in the oscillation formula: $|\mathsf{L}_{i0(kn)}|^2$ is $\mathcal{O}(n^{-2})$ and the sum up to $n$ gives $\mathcal{O}(n^{-1})$. In our study, for $|\bar{c}_i|\in[2.5,10]$, we choose the cutoff 40 to make the balance between computing time and accuracy.

\subsection{Experimental configurations}
\subsubsection{T2K experiment}
The Tokai-to-Kamioka (T2K) experiment is one of the two long-baseline neutrino oscillation facilities currently in operation. It employs a muon neutrino and anti-neutrino beam produced by a proton accelerator at J-PARC in Tokai, Japan, with a beam power of approximately 750~kW. The beam propagates 295 km to the Super-Kamiokande far detector (FD) in Kamioka.

The T2K neutrino beam is predominantly composed of muon neutrinos (96\%--98\%), with minor beam-related backgrounds, while the antineutrino beam similarly features a majority of muon antineutrinos. The dominant interaction mode for both neutrinos and antineutrinos is charged-current quasi-elastic (CCQE) scattering, supplemented by a smaller but significant contribution from resonant charged-current single-pion production (CC1$\pi$).

Data collected at Super-Kamiokande are typically separated into five samples: two appearance channels probing $\nu_\mu \to \nu_e$ ($\nu$-mode 1R$e$) and $\bar{\nu}_\mu \to \bar{\nu}_e$ ($\bar{\nu}$-mode 1R$e$) oscillations via CCQE interactions; two disappearance channels for $\nu_\mu \to \nu_\mu$ ($\nu$-mode 1R$\mu$) and $\bar{\nu}_\mu \to \bar{\nu}_\mu$ ($\bar{\nu}$-mode 1R$\mu$) via CCQE interactions; and one additional appearance channel dedicated to $\nu_\mu \to \nu_e$ oscillations through CC1$\pi$ interactions ($\nu$-mode 1R$e$1d$e$). In this work, we utilize the neutrino oscillation data accumulated during the run 1-10 of T2K from 2009 to 2020, as presented in Ref.~\cite{T2K:2023smv}.

\subsubsection{NOvA experiment}
The NuMI Off-axis $\nu_e$ Appearance (NOvA) experiment is the other currently operating long-baseline neutrino oscillation experiment. It produces muon neutrino and antineutrino beams using the NuMI beamline at Fermilab, Illinois, with an average beam power of around 700 kW. The neutrino/antineutrino beams travel 810 km underground to the NOvA far detector.

The neutrino and antineutrino data analyzed in this work are drawn from events recorded at the NOvA far detector between 2014 and 2024~\cite{Abubakar_2026}. These data are reported in nine samples: charged-current (CC) sub-samples for $\nu_\mu$ and $\bar{\nu}_\mu$; CC sub-samples for $\nu_e$ and $\bar{\nu}_e$ further divided by high/low event convolutional neural network (CNN) scores; $\nu_e$ samples with low enegry, and peripheral sub-samples.

 \subsection{Likelihood analysis}
 In this subsection we present the methodology of the likelihood analysis with $\chi^2$  in this study.
  We take the pull method~\cite{Huber:2002mx,Fogli:2002pt,GonzalezGarcia:2004wg} with the following formula ($\bar{\mu}_{1/3} $ means $\bar\mu_1$ for NH, $\bar\mu_3$ for IH):
\begin{equation}
\chi_{\text{pull}}^2(|\bar{c}|,\bar{\mu}_{1/3})=\underset{\vec\xi,\,\lambda^{\text{th}}_i}{\mathrm{min}}\left[\chi^2_{\text{stat}}(\vec\lambda,\vec\xi\,)+\sum_k\left(\frac{\xi_k}{\sigma_k}\right)^2+\sum_i\left(\frac{\lambda_i^{\text{th}}-\bar\lambda_i}{\sigma_i}\right)^2\right].
\label{chi2}
\end{equation}
In this formula, $\vec\lambda$ denotes all oscillation parameters (\{$\theta_{12}$, $\theta_{13}$, $\theta_{23}$, $\Delta m_{12}^2$, $\Delta m_{32}^2$, $\delta_{\text{CP}}$\} for standard oscillation and \{$\theta_{12}$, $\theta_{13}$, $\theta_{23}$, $\Delta m_{12}^2$, $\Delta m_{32}^2$, $\delta_{\text{CP}}$, $R$, $|\bar{c}|$, $\bar{\mu}_{1/3}$\} for DD model) and matter density $\rho$, the vector $\vec\xi$ denotes the systematic uncertainties, $\vec\lambda^{\rm{th}}$ denotes all parameters in $\vec\lambda$ except $|\bar{c}|$ and $\bar{\mu}_{1/3}$, and $\bar{\lambda}_i$ denotes the central values of $\lambda_i^{\text{th}}$, which will be explained in detail shortly. The minimization is performed over all systematic nuisance parameters $\vec\xi$ as well as some of the oscillation parameters $\lambda^{\rm{th}}_i$.

The statistical contribution is given by the Poissonian $\chi^2$:
\begin{equation}
    \chi^2_{\text{stat}}(\vec\lambda,\vec\xi\,)=2\sum_{\text{detectors}}\sum_{\text{channels}}\sum_{\text{bins}} \left(N^{\text{{th}}}-N^{\text{obs}}+N^{\text{obs}}\ln\frac{N^{\text{obs}}}{N^{\text{{th}}}} \right),
    \label{statistical}
\end{equation}
where $N^{\text{{th}}}$ and $N^{\text{obs}}$ are the sets of events corresponding to theoretical and observed data, respectively. The summation runs over detectors (the far detector of T2K or NOvA), channels (5 sample types for T2K and 9 for NOvA), and energy bins. 
 The second and third terms in Eq.~\eqref{chi2} account for the systematic and prior penalties. The systematic term constrains the nuisance parameters $\xi_k$ within their nominal uncertainties $\sigma_k$, while the prior term imposes a Gaussian penalty on deviations of values $\lambda_i^{\rm {th}}$ from their central values $\bar\lambda_i$, with $\sigma_i$ being the corresponding prior uncertainty of $1\sigma$. It is noted that central values and prior uncertainties of \{$\theta_{12}$, $\theta_{13}$, $\Delta m_{12}^2$\} are taken from global fit \cite{Esteban:2024eli} and which of \{$\theta_{23}$, $\Delta m_{32}^2$, $\delta_{\text{CP}}$\} are taken from experimental results \cite{T2K:2023smv,Abubakar_2026}. In this study, we consider four types of systematic uncertainty, with their values as follows: both T2K and NOvA adopt a 5\% uncertainty on signal normalization and a 10\% uncertainty on background normalization; for calibration uncertainties, T2K uses 0.01\% for both signal and background, while NOvA uses 2.5\% for both signal and background. 
\begin{figure}[!t]
    \centering
    \includegraphics[width=0.9\linewidth]{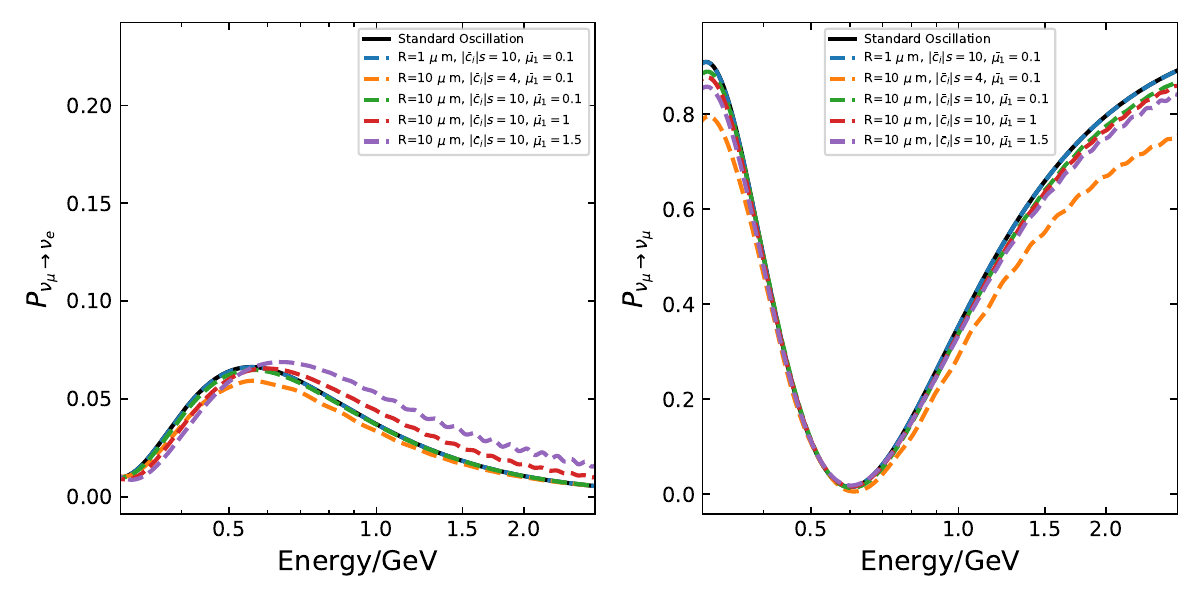}
    \includegraphics[width=0.9\linewidth]{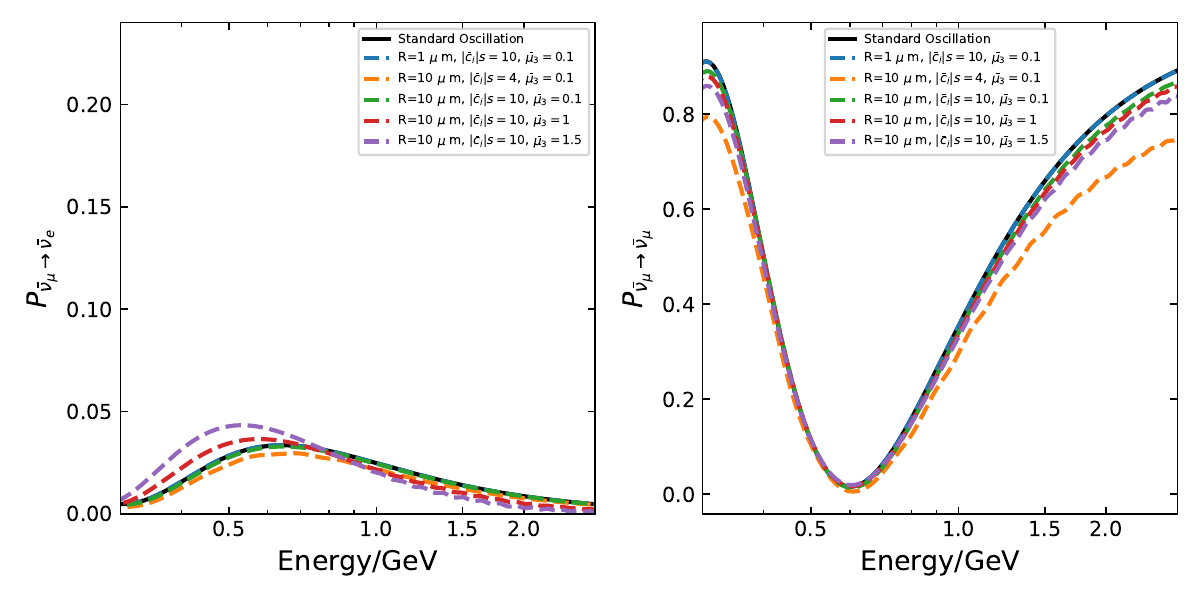}
    \caption{Neutrino oscillation probabilities for both standard oscillation and DD models at the FD with a baseline of 295 km in T2K. The top (bottom) row refers to the case of neutrinos (anti-neutrinos) while the two columns show the cases of the two channels $\nu_\mu\to\nu_e$ and $\nu_\mu\to\nu_\mu$. The mass hierarchy is considered as NH. The black curves correspond to the standard oscillation case, while the other curves show the presence of the Dark Dimension with different DD parameter settings, as shown in the figure legends. }
    \label{T2Kprob}
\end{figure}

For a given set of the central values $\bar\lambda_i$, the estimate $\chi^2$ as a function of the parameters of interest (namely the DD parameters $|\bar{c}_i|$ and $\bar{\mu}_1$/$\bar{\mu}_3$ (NH/IH)) is obtained by minimizing the expression inside the square brackets in Eq.~\eqref{chi2} over a set of parameters 
out of $\vec\lambda^{\text{th}}$ and all systematic nuisance parameters, while keeping the remaining parameters fixed. This is equivalent to the minimization over $\vec\lambda^{\text{th}}$ and $\vec\xi$ with the uncertainties $\sigma_i$ set to zero for the parameters in $\vec\lambda^{\text{th}}$ that we fix. In this work, we will consider two approaches: one involves this minimization procedure over the parameter set \{$\theta_{13}$, $\Delta m_{32}^2$, $\vec\xi$\,\} (commonly referred to as the marginalization over  \{$\theta_{13}$, $\Delta m_{32}^2$; $\vec\xi$\} \cite{Huber:2004ka,Huber:2006xx,Siyeon:2024pte}), and the other keeps all oscillation parameters fixed without marginalization.

\begin{figure}[!t]
    \centering
    \includegraphics[width=0.9\linewidth]{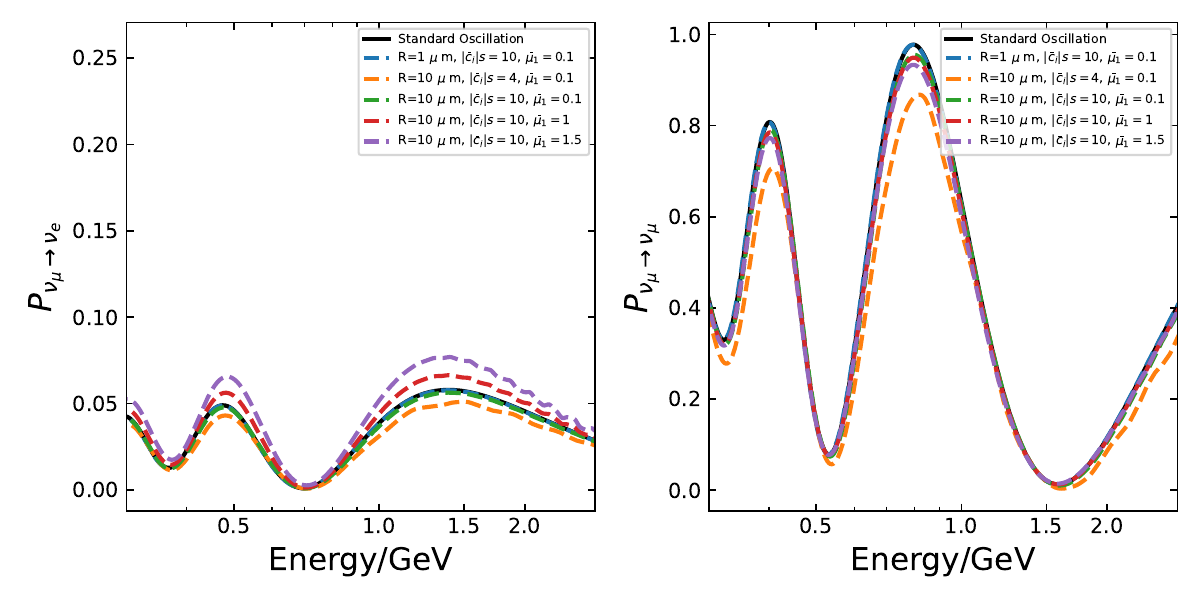}
    \includegraphics[width=0.9\linewidth]{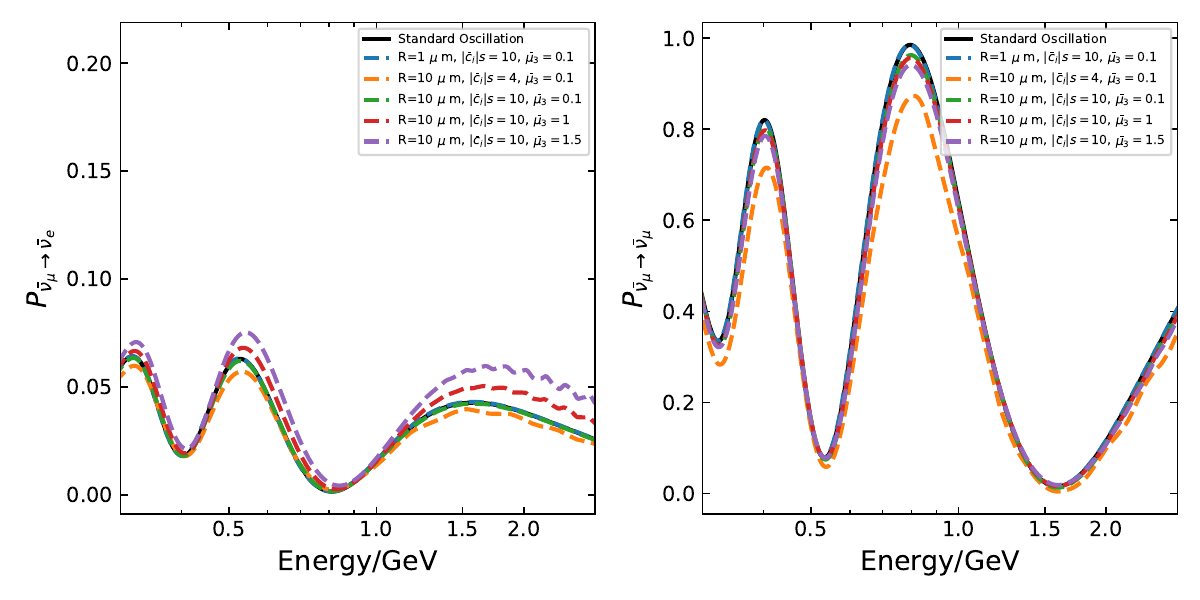}
    \caption{Oscillation probabilities for both standard oscillation and DD models at the NOvA FD baseline of 810 km. The top (bottom) row refers to the case
 of neutrinos (anti-neutrinos) while
  the two columns show the cases
 of the two channels $\nu_\mu\to\nu_e$ and $\nu_\mu\to\nu_\mu$. The black curves correspond to the 4D standard oscillation scenario, while the other curves manifest deviations from it due to extra direction with different model parameter sets. The mass hierarchy is chosen to be NH.}
    \label{NOvAprob}
\end{figure}

\subsection{Numerical results}
\subsubsection{Neutrino oscillation probabilities in presence of the Dark Dimension}
\label{prob}

Figures~\ref{T2Kprob} and~\ref{NOvAprob} show the oscillation probabilities under the standard oscillation scenario and our DD model, with the mass hierarchy taken as NH and the baselines set to match the T2K and NOvA experiments, respectively. The black solid lines represent the standard oscillation scenario, while the dashed lines correspond to the DD model with different parameter settings. The standard oscillation parameters and DD parameter settings are detailed in Section~\ref{settings}.

 We note that the presence of an extra dimension introduces new frequencies on top of those from the active neutrino mass square differences, leading to small sawtooth-like fluctuations in the probability curves. We therefore applied a low-pass filter (0.03~GeV) to both standard oscillation and DD scenarios in all channels. The blue dashed lines in Figures~\ref{T2Kprob} and~\ref{NOvAprob} correspond to the DD parameters \{$R$=1\,$\mu$m, $|\bar{c}_i|=10$, $\bar{\mu}_1=0.1$\}. In this setup, the mixing coefficients for higher modes are very small and the DD model degenerates into the standard oscillation. It can be seen that it closely follows the standard oscillation curve, indicating that the filter does not introduce additional differences in the probabilities.

\begin{figure}[!t]
    \centering
    \includegraphics[width=1\linewidth]{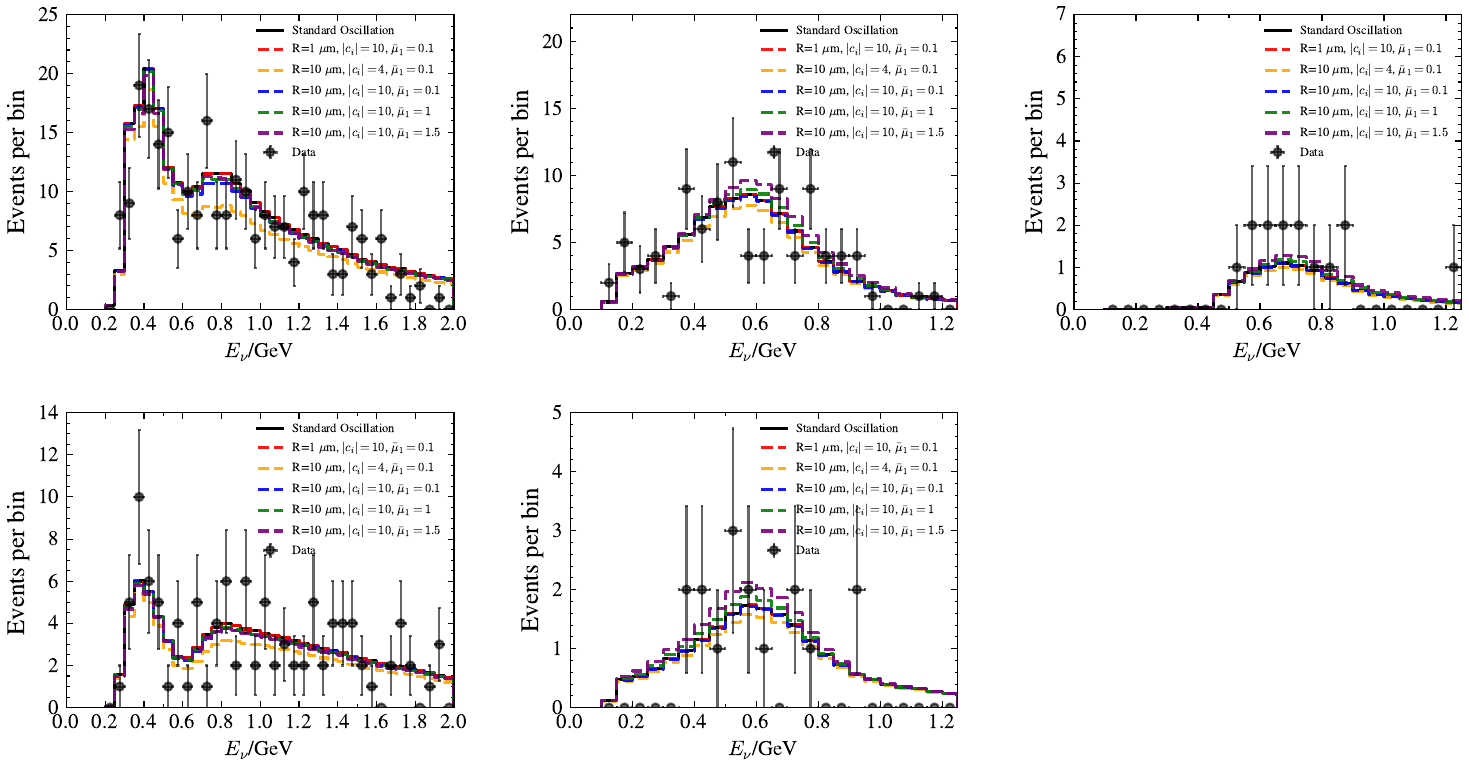}
    \caption{The reconstructed neutrino energy spectra for the FD
$\nu_\mu$ CC and $\nu_e$ CC samples for T2K. The five subfigures refer to five channels of T2K: $\nu$-mode 1R$\mu$, $\nu$-mode 1R$e$, $\nu$-mode 1R$e$1d$e$, $\bar{\nu}$-mode 1R$\mu$ and $\bar{\nu}$-mode 1R$e$, whose order is the same as the order in dataset \cite{T2K:2023smv}. The solid curves correspond to the standard oscillation, while the dashed curves correspond to the DD case with different 
 sets of parameters, which are shown in the figure legends.}
    \label{T2Kevent}
\end{figure}
 
For the other DD parameter settings shown in the figures, the impact of the DD model on the oscillation probabilities becomes evident. In general, larger values of $R$ amplify the deviations from standard oscillations across all channels in both experiments. The small $|\bar{c}_i|$ (orange dashed line) suppresses the oscillation probabilities of all channels. The large $\bar{\mu}_1$ (purple dashed line) decreases the oscillation probability for $\nu_\mu\to\nu_\mu$; for T2K, this parameter setting increases the oscillation probability for $\bar{\nu}_\mu\to\bar{\nu}_e$ at low energies and for $\nu_\mu\to\nu_e$ at high energies, while for NOvA it increases  $P_{\nu_\mu\to \nu_e}$ and $P_{\bar{\nu}_\mu\to \bar{\nu}_e}$ across all energies. For larger $|\bar{c}_i|$ and smaller $\bar{\mu}_1$ (green and red dashed lines), the oscillation probabilities return to the standard scenario. Additionally, for larger $\bar{\mu}_1$, small sawtooth-like fluctuations appear in the probability curves, which are residual rapid oscillations in the DD model after filtering.

\subsubsection{Neutrino oscillation spectra with Dark Dimension models}
\begin{figure}[!t]
    \centering
    \includegraphics[width=0.95\linewidth]{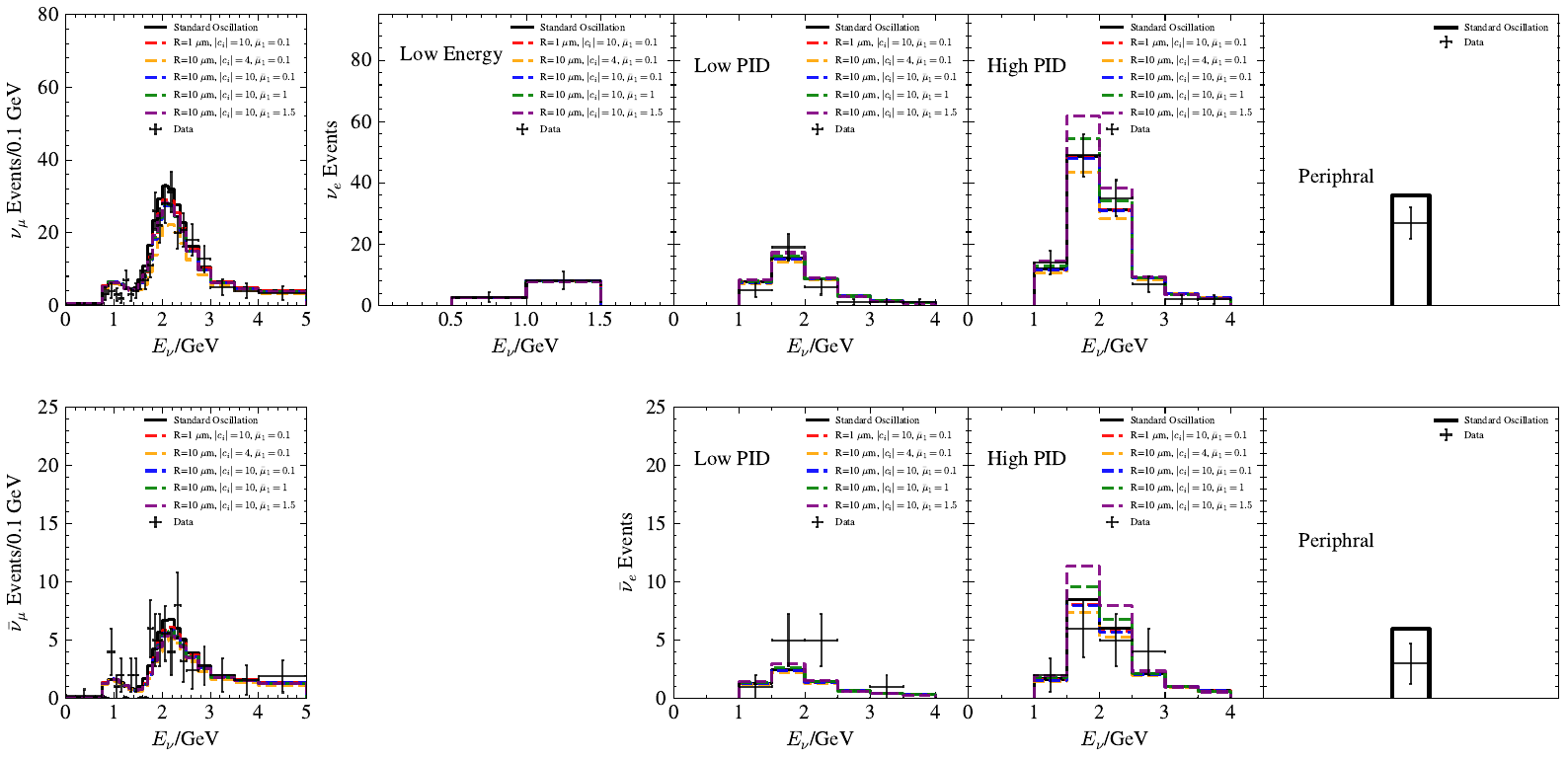}
    \caption{The reconstructed neutrino energy spectra for the FD
$\nu_\mu$ and $\nu_e$ samples for NOvA. The
 top (bottom) row corresponds to the neutrino (anti-neutrino) channel. The five columns above correspond to
 the $\nu_\mu$-like and $\nu_e$-like events with low energy, high CNN, low CNN, and peripheral, while the four columns below correspond to
 the $\bar\nu_\mu$-like and $\bar\nu_e$-like events with high CNN, low CNN, and peripheral, respectively. Their order is consistent with the order in the dataset \cite{Abubakar_2026}. The solid curves correspond to the standard oscillation, while the dashed curves correspond to the DD case with different 
 sets of parameters, which are shown in the figure legends. }
    \label{NOvAspectra}
\end{figure}

We present the reconstructed neutrino energy spectra for T2K and NOvA, as shown in Figures~\ref{T2Kevent} and~\ref{NOvAspectra}. For T2K, we study the DD model in five channels: $\nu$-mode 1R$\mu$, $\nu$-mode 1R$e$, $\nu$-mode 1R$e$1d$e$, $\bar{\nu}$-mode 1R$\mu$, and $\bar{\nu}$-mode 1R$e$. For NOvA, we investigate the DD model in seven channels: $\nu_\mu$-like channel, $\nu_e$-like events with high/low CNN score, the corresponding anti-neutrino channels, and $\nu_e$-like events with low energy. The energies of peripheral neutrino and anti-neutrino channels of NOvA are not reported in the datasets, so we show their event rates only for completeness. The standard oscillation parameters and DD parameter settings are given in Section~\ref{settings}.

Following the analysis of oscillation probabilities, we first examine the case where the DD parameters are set so that the model degenerates to the standard oscillation. It can be seen that the red dashed line is highly overlapped with the standard oscillation (black solid line).

In light of different DD parameter settings, the deviations from the standard oscillation scenario are clearly visible. In particular, small $|\bar{c}_i|$ (yellow dashed lines) reduces the event rates in all channels. The larger value of $\bar{\mu}_i$ slightly decreases the event rates in the $\nu_\mu$ and $\bar{\nu}_\mu$ channels (purple dashed lines) while increasing the event rates in the $\nu_e$ and $\bar{\nu}_e$ channels. On the other hand, for larger $|\bar{c}_i|$ and smaller $\bar{\mu}_i$ (blue and green dashed lines), the event rates approach the standard oscillation scenario, as expected.

\subsubsection{Constraints on DD model parameters}

\begin{figure}[!t]
    \centering    
    \includegraphics[width=\textwidth]{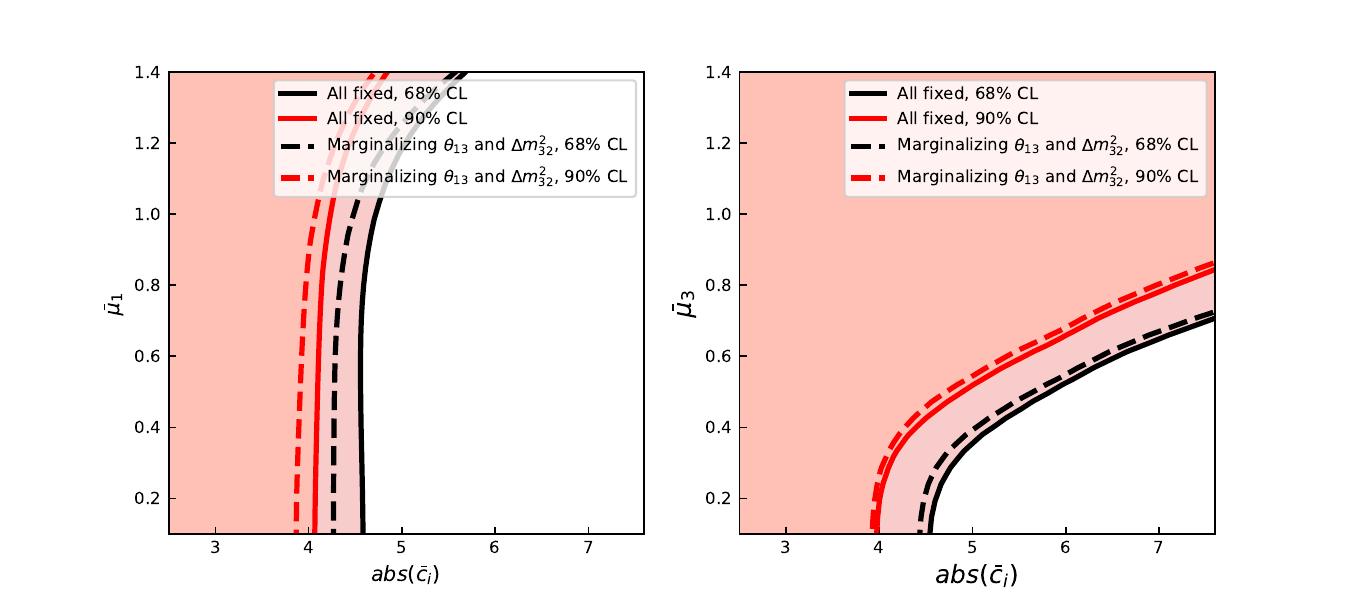}
    \caption{The exclusion limits at 68\% (black line) and 90\% (red line) C.L. in the T2K experiment. The solid and dashed lines distinguish two marginalization setups: the solid lines are obtained by fixing all standard oscillation parameters for NH and IH, while the dashed lines are obtained by marginalizing $\theta_{13}$ and $\Delta m_{32}^2$. The DD parameters in the shaded region are excluded. }
    \label{T2K_Scan}
\end{figure}

\begin{figure}[htb]
    \centering
    \includegraphics[width=\linewidth]{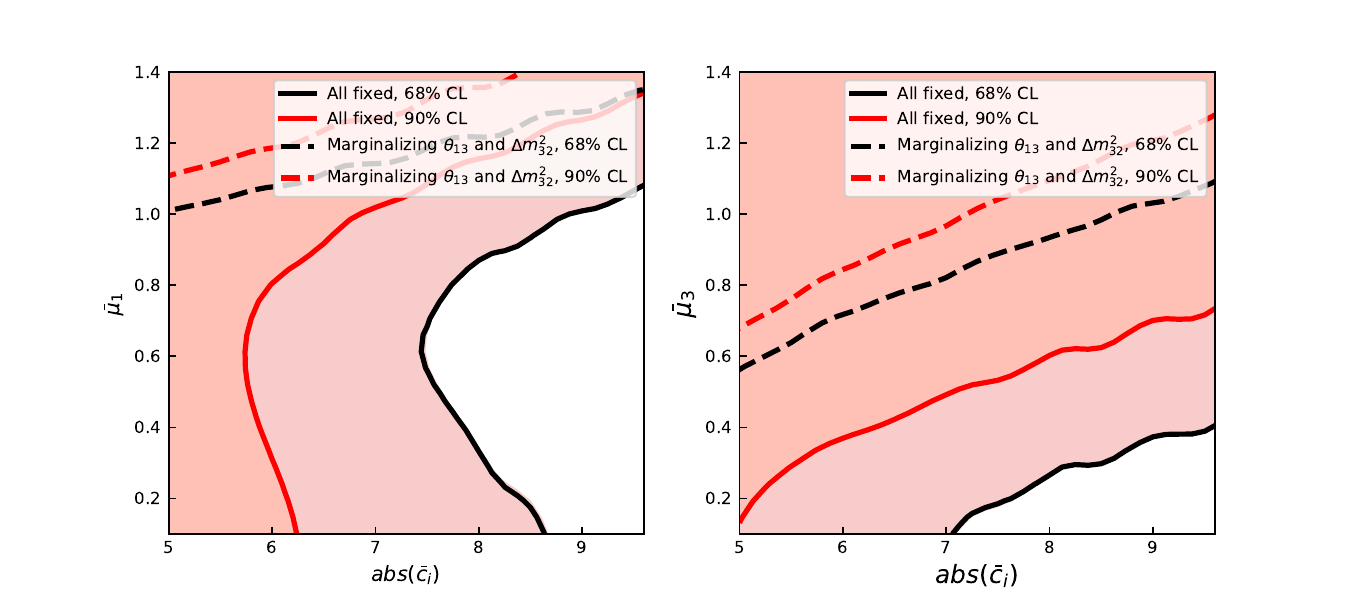}
    \caption{The exclusion limits at 68\% (black line) and 90\% (red line) C.L. in the NOvA experiment. The solid and dashed lines distinguish two marginalization setups: the solid lines are obtained by fixing all standard oscillation parameters for NH and IH, while the dashed line are obtained by marginalizing $\theta_{13}$ and $\Delta m_{32}^2$. The DD parameters in the shaded region are excluded.}
    \label{NOvA_Scan}
\end{figure}

After computing $\chi^2$ for the standard oscillation and the DD model, we obtain the contours of $\Delta\chi^2$ at 68\% and 90\% C.L. from the T2K and NOvA experiments in the parameter space of $|\bar{c}_i|$--$\bar{\mu}_1$ (NH) or $|\bar{c}_i|$--$\bar{\mu}_3$ (IH), as shown in Figures~\ref{T2K_Scan} and~\ref{NOvA_Scan}. We consider two different marginalization settings: fixing all standard oscillation parameters and marginalizing over $\Delta m_{32}^2$ and $\theta_{13}$. For the DD parameters, $|\bar{c}_i|$ and $\bar{\mu}_1$/$\bar{\mu}_3$ (NH/IH) vary, while $R$ is fixed at 10~$\mu$m to produce sufficient differences. The remaining DD parameters $\bar{\mu}_2$, $\bar{\mu}_3$ (NH) or $\bar{\mu}_1$, $\bar{\mu}_2$ (IH), as well as the standard oscillation parameters, are taken as described in Section~\ref{settings}.


In all figures, it can be seen that regions with smaller $|\bar{c}_i|$ and larger $|\bar{\mu}_i|$ are excluded, since such settings would increase the mixing coefficients and introduce a significant difference compared to the standard neutrino oscillation. In addition, when $\bar{\mu}_i$ is large, the small oscillations in the exclusion curves originate from secondary oscillations in the probability curves.

Regarding the T2K experiment, the exclusion curves for NH are independent of $\bar{\mu}_1$ at small $\bar{\mu}_1$, while bending towards larger $|\bar{c}_i|$ at larger $\bar{\mu}_1$. For IH, the curves bend toward larger $|\bar{c}_i|$ even at small $\bar{\mu}_3$ values. 
In contrast, the exclusion curves of the NOvA experiment show a less significant dependence on $\bar{\mu}_1$ ($\bar{\mu}_3$) at lower values for NH (IH). 
For both mass hierarchies and marginalization strategies, the excluded regions of the NOvA experiment are larger than those of the T2K experiment.
One can also see that when $\theta_{13}$ and $\Delta m_{32}^2$ are marginalized over, the exclusion contours shift toward regions with larger $\bar{\mu}_i$ and smaller $|\bar{c}_i|$, resulting in smaller excluded parameter regions, while keeping the overall shape of the curves.

These exclusion plots suggest lower bounds on the bulk mass of the R-neutrinos. For example, for the DD model with $R=10~\mu\text{m}$ and $\bar\mu_{1/3} = 0.1$, we find a lower bound $|\bar{c}_1| \gtrsim 6.2$ for NH and $|\bar{c}_3| \gtrsim 4.9$ for IH.

\section{Summary and Outlook}

In this article, we have studied the properties of right-handed neutrinos propagating along an extra dimension of radius $ R =10~\mu\text{m}$ motivated by the DD scenario, with the bulk masses $|c_i|$  larger than the compactification scale  $R^{-1}$ or equivalently the corresponding dimensionless parameters $|\bar c_i| = |c_i R| > 1$. We have demonstrated that, at a high confidence level, long-baseline neutrino oscillation experiments have a strict exclusion limit on the extra dimension model parameters. The derived constraints on dark dimension right-handed neutrinos are complementary to those results from the collider experiments and cosmological observations. 

Using data from T2K and NOvA experiments, we showed that there is a lower bound for the bulk mass in the DD scenario. Fixing $R=10~\mu\text{m}$, the lower bound is $|\bar c_1| \gtrsim 6.2$ or equivalently $|c_1| \gtrsim 0.12~\text{eV}$. In IH, we get a smaller lower bound $|\bar c_3| \gtrsim 4.9$ or, equivalently, $|c_3| \gtrsim 0.10~\text{eV}$. For a DD scenario with smaller radius, including the two dark dimension proposal~\cite{Anchordoqui:2025nmb} with sub-micrometer size, these bounds could be much lower. 

It should be mentioned that for $|\bar c_i| \lesssim 1$, the first KK mode dominates the transition probability. In \cite{Carena:2017qhd}, it was suggested that a particular scenario of the 3+1 sterile neutrino model would produce a result that is roughly equivalent to the extra dimension model. For $|\bar c_i| > 1$, this approximate correspondence with a 3+1 model appears to be lost as the first KK mode is suppressed by the bulk mass, and the $|\bar c_i|$-th excitation with a mass of order the bulk mass becomes dominant. A more accurate approximation of the transition probability requires a large number of KK modes. However, as demonstrated in \cite{Antoniadis:2025rck}, when the bulk mass is sufficiently large compared to the compactification scale, there exists another 3+1 limit where there is a quasi-continuum of KK excitations above the bulk mass that acts as a mass gap above very light active neutrino zero modes. This 3+1 limit has been explored in the region $25<|\bar c_i|<2000$ using data from the KATRIN's eV sterile neutrino search \cite{KATRIN:2025lph}. For future projects, it would be interesting to extend the analysis in this paper to cover a larger parameter region. For example, the data from the reactor neutrino experiments would be useful to study the parameter region with very large $|\bar c_i|$. 
Following the same strategy, it is promising for future long-baseline neutrino oscillation experiments such as T2HK and DUNE to discover or exclude the dark dimension right-handed neutrinos. To have a more complete picture, it will also be worthwhile in a future work to combine our analysis  with data from other existing atmospheric neutrino experiments such as Super-Kamiokande and IceCube~\cite{Super-Kamiokande:2023ahc,T2K:2024wfn,IceCubeCollaboration:2024ssx}.

An interesting extension of this work would be to relax the real-valued bulk-brane coupling constants to complex-valued ones, which will lead to CP violation from KK modes. This direction as well as its consequences in particle physics and cosmology would be worth investigating.

\section*{Acknowledgments}
 This project is supported in part by National Natural Science Foundation of China under Grant No.~12347105 and Fundamental Research Funds for the Central Universities (23xkjc017), Sun Yat-Sen University. The work of AC is funded by the NSRF via the Program Management Unit for Human Resources $\&$ Institutional Development, Research and Innovation (grant B39G680010). HI is supported by the Second Century Fund (C2F), Chulalongkorn University. We are grateful to Ignatios Antoniadis, Kento Asai, Mitesh Kumar Behera, Chakrit Pongkitivanichkul,  Norraphat Srimanobhas and Nakorn Thongyoi for valuable discussions. 

\bibliographystyle{JHEP}
\bibliography{neutrino}

@article{Siyeon:2024pte,
    author = "Siyeon, Kim and Kim, Suhyeon and Masud, Mehedi and Park, Juseong",
    title = "{Probing large extra dimension at DUNE using beam tunes}",
    eprint = "2409.08620",
    archivePrefix = "arXiv",
    primaryClass = "hep-ph",
    doi = "10.1007/JHEP11(2024)141",
    journal = "JHEP",
    volume = "11",
    pages = "141",
    year = "2024"
}

@article{deGiorgi:2025xgp,
    author = "de Giorgi, Arturo and Pasari, Dhruv and Turner, Jessica",
    title = "{Do neutrinos dream in 5D? Towards a comprehensive extra-dimensional neutrino phenomenology}",
    eprint = "2512.02101",
    archivePrefix = "arXiv",
    primaryClass = "hep-ph",
    reportNumber = "IPPP/25/84, IPPP/25/84",
    doi = "10.1007/JHEP05(2026)152",
    journal = "JHEP",
    volume = "05",
    pages = "152",
    year = "2026"
}

@article{Elacmaz:2025ihm,
    author = {Ela{\c{c}}maz, T. G{\"o}kalp and Martinez-Soler, Ivan and Perez-Gonzalez, Yuber F.},
    title = "{Updated Constraints on Large Extra Dimensions from Reactor Antineutrino Experiments}",
    eprint = "2510.12900",
    archivePrefix = "arXiv",
    primaryClass = "hep-ph",
    reportNumber = "IFT-UAM/CSIC-25-109, IPPP/25/66",
    month = "10",
    year = "2025"
}

@article{Panda:2024ioo,
    author = "Panda, Papia and Mishra, Priya and Roy, Samiran and Ghosh, Monojit and Mohanta, Rukmani",
    title = "{Study of Large Extra Dimension and neutrino decay at P2SO experiment}",
    eprint = "2411.09628",
    archivePrefix = "arXiv",
    primaryClass = "hep-ph",
    doi = "10.1007/JHEP05(2025)018",
    journal = "JHEP",
    volume = "05",
    pages = "018",
    year = "2025"
}

@article{Dvali:1999cn,
    author = "Dvali, G. R. and Smirnov, Alexei Yu.",
    title = "{Probing large extra dimensions with neutrinos}",
    eprint = "hep-ph/9904211",
    archivePrefix = "arXiv",
    reportNumber = "NYU-TH-99-3-03",
    doi = "10.1016/S0550-3213(99)00574-X",
    journal = "Nucl. Phys. B",
    volume = "563",
    pages = "63--81",
    year = "1999"
}

@article{Lukas:2000wn,
    author = "Lukas, Andre and Ramond, Pierre and Romanino, Andrea and Ross, Graham G.",
    title = "{Solar neutrino oscillation from large extra dimensions}",
    eprint = "hep-ph/0008049",
    archivePrefix = "arXiv",
    reportNumber = "OUTP-00-34P",
    doi = "10.1016/S0370-2693(00)01206-5",
    journal = "Phys. Lett. B",
    volume = "495",
    pages = "136--146",
    year = "2000"
}

@article{Machado:2011jt,
    author = "Machado, P. A. N. and Nunokawa, H. and Zukanovich Funchal, R.",
    title = "{Testing for Large Extra Dimensions with Neutrino Oscillations}",
    eprint = "1101.0003",
    archivePrefix = "arXiv",
    primaryClass = "hep-ph",
    doi = "10.1103/PhysRevD.84.013003",
    journal = "Phys. Rev. D",
    volume = "84",
    pages = "013003",
    year = "2011"
}

@article{Carena:2017qhd,
    author = "Carena, Marcela and Li, Ying-Ying and Machado, Camila S. and Machado, Pedro A. N. and Wagner, Carlos E. M.",
    title = "{Neutrinos in Large Extra Dimensions and Short-Baseline $\nu_e$ Appearance}",
    eprint = "1708.09548",
    archivePrefix = "arXiv",
    primaryClass = "hep-ph",
    reportNumber = "FERMILAB-PUB-17-338-T",
    doi = "10.1103/PhysRevD.96.095014",
    journal = "Phys. Rev. D",
    volume = "96",
    number = "9",
    pages = "095014",
    year = "2017"
}

@article{Forero:2022skg,
    author = "Forero, D. V. and Giunti, C. and Ternes, C. A. and Tyagi, O.",
    title = "{Large extra dimensions and neutrino experiments}",
    eprint = "2207.02790",
    archivePrefix = "arXiv",
    primaryClass = "hep-ph",
    doi = "10.1103/PhysRevD.106.035027",
    journal = "Phys. Rev. D",
    volume = "106",
    number = "3",
    pages = "035027",
    year = "2022"
}

@article{Eller:2025lsh,
    author = "Eller, Philipp and Ettengruber, Manuel and Zander, Alan",
    title = "{Neutrino data analysis of extra-dimensional theories with massive bulk fields}",
    eprint = "2508.04274",
    archivePrefix = "arXiv",
    primaryClass = "hep-ph",
    doi = "10.1103/1llm-96vy",
    journal = "Phys. Rev. D",
    volume = "112",
    number = "5",
    pages = "055009",
    year = "2025"
}

@article{T2K:2023smv,
    author = "Abe, K. and others",
    collaboration = "T2K",
    title = "{Measurements of neutrino oscillation parameters from the T2K experiment using $3.6\times 10^{21}$ protons on target}",
    eprint = "2303.03222",
    archivePrefix = "arXiv",
    primaryClass = "hep-ex",
    doi = "10.1140/epjc/s10052-023-11819-x",
    journal = "Eur. Phys. J. C",
    volume = "83",
    number = "9",
    pages = "782",
    year = "2023"
}

@article{Antoniadis:2025rck,
    author = "Antoniadis, Ignatios and Chatrabhuti, Auttakit and Isono, Hiroshi",
    title = "{Searching for a dark dimension right-handed neutrino in KATRIN}",
    eprint = "2509.05233",
    archivePrefix = "arXiv",
    primaryClass = "hep-ph",
    doi = "10.1007/JHEP02(2026)015",
    journal = "JHEP",
    volume = "02",
    pages = "015",
    year = "2026"
}

@article{Planck:2018nkj,
    author = "Aghanim, N. and others",
    collaboration = "Planck",
    title = "{Planck 2018 results. I. Overview and the cosmological legacy of Planck}",
    eprint = "1807.06205",
    archivePrefix = "arXiv",
    primaryClass = "astro-ph.CO",
    doi = "10.1051/0004-6361/201833880",
    journal = "Astron. Astrophys.",
    volume = "641",
    pages = "A1",
    year = "2020"
}

@article{Planck:2018vyg,
    author = "Aghanim, N. and others",
    collaboration = "Planck",
    title = "{Planck 2018 results. VI. Cosmological parameters}",
    eprint = "1807.06209",
    archivePrefix = "arXiv",
    primaryClass = "astro-ph.CO",
    doi = "10.1051/0004-6361/201833910",
    journal = "Astron. Astrophys.",
    volume = "641",
    pages = "A6",
    year = "2020",
    note = "[Erratum: Astron.Astrophys. 652, C4 (2021)]"
}

@article{Montero:2022prj,
    author = "Montero, Miguel and Vafa, Cumrun and Valenzuela, Irene",
    title = "{The dark dimension and the Swampland}",
    eprint = "2205.12293",
    archivePrefix = "arXiv",
    primaryClass = "hep-th",
    doi = "10.1007/JHEP02(2023)022",
    journal = "JHEP",
    volume = "02",
    pages = "022",
    year = "2023"
}

@article{Anchordoqui:2022svl,
    author = "Anchordoqui, Luis A. and Antoniadis, Ignatios and Lust, Dieter",
    title = "{Aspects of the dark dimension in cosmology}",
    eprint = "2212.08527",
    archivePrefix = "arXiv",
    primaryClass = "hep-ph",
    reportNumber = "MPP-2022-285, LMU-ASC 55/22",
    doi = "10.1103/PhysRevD.107.083530",
    journal = "Phys. Rev. D",
    volume = "107",
    number = "8",
    pages = "083530",
    year = "2023"
}

@article{Antoniadis:1990ew,
    author = "Antoniadis, Ignatios",
    title = "{A Possible new dimension at a few TeV}",
    reportNumber = "EP-CPTH-A978-0690",
    doi = "10.1016/0370-2693(90)90617-F",
    journal = "Phys. Lett. B",
    volume = "246",
    pages = "377--384",
    year = "1990"
}

@article{Arkani-Hamed:1998jmv,
    author = "Arkani-Hamed, Nima and Dimopoulos, Savas and Dvali, G. R.",
    title = "{The Hierarchy problem and new dimensions at a millimeter}",
    eprint = "hep-ph/9803315",
    archivePrefix = "arXiv",
    reportNumber = "SLAC-PUB-7769, SU-ITP-98-13",
    doi = "10.1016/S0370-2693(98)00466-3",
    journal = "Phys. Lett. B",
    volume = "429",
    pages = "263--272",
    year = "1998"
}

@article{Antoniadis:1998ig,
    author = "Antoniadis, Ignatios and Arkani-Hamed, Nima and Dimopoulos, Savas and Dvali, G. R.",
    title = "{New dimensions at a millimeter to a Fermi and superstrings at a TeV}",
    eprint = "hep-ph/9804398",
    archivePrefix = "arXiv",
    reportNumber = "SLAC-PUB-7801, SU-ITP-98-28, CPTH-S608-0498, IC-98-39",
    doi = "10.1016/S0370-2693(98)00860-0",
    journal = "Phys. Lett. B",
    volume = "436",
    pages = "257--263",
    year = "1998"
}

@article{Dienes:1998sb,
    author = "Dienes, Keith R. and Dudas, Emilian and Gherghetta, Tony",
    title = "{Neutrino oscillations without neutrino masses or heavy mass scales: A Higher dimensional seesaw mechanism}",
    eprint = "hep-ph/9811428",
    archivePrefix = "arXiv",
    reportNumber = "CERN-TH-98-370",
    doi = "10.1016/S0550-3213(99)00377-6",
    journal = "Nucl. Phys. B",
    volume = "557",
    pages = "25",
    year = "1999"
}

@article{Arkani-Hamed:1998wuz,
    author = "Arkani-Hamed, Nima and Dimopoulos, Savas and Dvali, G. R. and March-Russell, John",
    title = "{Neutrino masses from large extra dimensions}",
    eprint = "hep-ph/9811448",
    archivePrefix = "arXiv",
    reportNumber = "SLAC-PUB-8014, SU-ITP-98-64",
    doi = "10.1103/PhysRevD.65.024032",
    journal = "Phys. Rev. D",
    volume = "65",
    pages = "024032",
    year = "2001"
}

@article{Ooguri:2006in,
    author = "Ooguri, Hirosi and Vafa, Cumrun",
    title = "{On the Geometry of the String Landscape and the Swampland}",
    eprint = "hep-th/0605264",
    archivePrefix = "arXiv",
    reportNumber = "CALT-68-2600, HUTP-06-A017",
    doi = "10.1016/j.nuclphysb.2006.10.033",
    journal = "Nucl. Phys. B",
    volume = "766",
    pages = "21--33",
    year = "2007"
}

@article{Lust:2019zwm,
    author = {L{\"u}st, Dieter and Palti, Eran and Vafa, Cumrun},
    title = "{AdS and the Swampland}",
    eprint = "1906.05225",
    archivePrefix = "arXiv",
    primaryClass = "hep-th",
    doi = "10.1016/j.physletb.2019.134867",
    journal = "Phys. Lett. B",
    volume = "797",
    pages = "134867",
    year = "2019"
}

@article{Anchordoqui:2025nmb,
    author = "Anchordoqui, Luis A. and Antoniadis, Ignatios and Lust, Dieter",
    title = "{Two Micron-Size Dark Dimensions}",
    eprint = "2501.11690",
    archivePrefix = "arXiv",
    primaryClass = "hep-th",
    reportNumber = "MPP-2025-5, LMU-ASC 02/25",
    doi = "10.1002/prop.70015",
    journal = "Fortsch. Phys.",
    volume = "73",
    number = "8",
    pages = "e70015",
    year = "2025"
}

@article{Anchordoqui:2023wkm,
    author = "Anchordoqui, Luis A. and Antoniadis, Ignatios and Cunat, Jules",
    title = "{Dark dimension and the standard model landscape}",
    eprint = "2306.16491",
    archivePrefix = "arXiv",
    primaryClass = "hep-ph",
    doi = "10.1103/PhysRevD.109.016028",
    journal = "Phys. Rev. D",
    volume = "109",
    number = "1",
    pages = "016028",
    year = "2024"
}

@article{KATRIN:2025lph,
    author = "Acharya, Himal and others",
    collaboration = "KATRIN",
    title = "{Sterile-neutrino search based on 259 days of KATRIN data}",
    eprint = "2503.18667",
    archivePrefix = "arXiv",
    primaryClass = "hep-ex",
    doi = "10.1038/s41586-025-09739-9",
    journal = "Nature",
    volume = "648",
    number = "8092",
    pages = "70--75",
    year = "2025"
}

@article{Roy:2023dyq,
    author = "Roy, Samiran",
    title = "{Capability of the proposed long-baseline experiments to probe large extra dimension}",
    eprint = "2305.16234",
    archivePrefix = "arXiv",
    primaryClass = "hep-ph",
    doi = "10.1103/PhysRevD.108.055015",
    journal = "Phys. Rev. D",
    volume = "108",
    number = "5",
    pages = "055015",
    year = "2023"
}

@article{Lukas:2000rg,
    author = "Lukas, Andre and Ramond, Pierre and Romanino, Andrea and Ross, Graham G.",
    title = "{Neutrino Masses and Mixing in Brane World Theories}",
    eprint = "hep-ph/0011295",
    archivePrefix = "arXiv",
    reportNumber = "FERMILAB-PUB-00-284-T, OUTP-00-39P, SUSX-TH-00-018, UFIFT-HET-00-27",
    doi = "10.1088/1126-6708/2001/04/010",
    journal = "JHEP",
    volume = "04",
    pages = "010",
    year = "2001"
}

@article{Esteban:2024eli,
    author = "Esteban, Ivan and Gonzalez-Garcia, M. C. and Maltoni, Michele and Martinez-Soler, Ivan and Pinheiro, Jo{\~a}o Paulo and Schwetz, Thomas",
    title = "{NuFit-6.0: updated global analysis of three-flavor neutrino oscillations}",
    eprint = "2410.05380",
    archivePrefix = "arXiv",
    primaryClass = "hep-ph",
    reportNumber = "IFT-UAM/CSIC-24-140, YITP-SB-2024-24, IPPP/24/64, IPPP/24/64, IFT-UAM/CSIC-24-140, YITP-SB-2024-24",
    doi = "10.1007/JHEP12(2024)216",
    journal = "JHEP",
    volume = "12",
    pages = "216",
    year = "2024"
}

@article{Huber:2004ka,
    author = "Huber, Patrick and Lindner, M. and Winter, W.",
    title = "{Simulation of long-baseline neutrino oscillation experiments with GLoBES (General Long Baseline Experiment Simulator)}",
    eprint = "hep-ph/0407333",
    archivePrefix = "arXiv",
    reportNumber = "TUM-HEP-553-04",
    doi = "10.1016/j.cpc.2005.01.003",
    journal = "Comput. Phys. Commun.",
    volume = "167",
    pages = "195",
    year = "2005"
}

@article{Huber:2006xx,
    author         = "Huber, Patrick and Kopp, Joachim and Lindner, Manfred
                       and Rolinec, Mark and Winter, Walter",
    title          = "{New features in the simulation of neutrino oscillation
                       experiments with GLoBES 3.0: General long baseline experiment simulator}",
    journal        = "Comput. Phys. Commun.",
    volume         = "177",
    year           = "2007",
    pages          = "432-438",
    doi            = "10.1016/j.cpc.2007.01.001",
    eprint         = "hep-ph/0601266",
    archivePrefix  = "arXiv",
    primaryClass   = "hep-ph",
    SLACcitation   = "%%CITATION = HEP-PH/0601266;%%"
}

@article{Huber:2002mx,
  author    = {Patrick Huber and Manfred Lindner and Walter Winter},
  title     = {Superbeams versus Neutrino Factories},
  journal   = {Nucl.Phys.},
  volume    = {B645},
  pages     = {3-48},
  year      = {2002},
  doi       = {10.1016/S0550-3213(02)00825-8},
  eprint    = {hep-ph/0204352},
  archivePrefix = {arXiv},
  primaryClass = {hep-ph}
}

@article{Fogli:2002pt,
  author        = {G.L. Fogli and E. Lisi and A. Marrone and D. Montanino and A. Palazzo},
  title         = {Getting the most from the statistical analysis of solar neutrino oscillations},
  journal       = {Phys. Rev. D},
  volume        = {66},
  pages         = {053010},
  year          = {2002},
  doi           = {10.1103/PhysRevD.66.053010},
  eprint        = {hep-ph/0206162},
  archivePrefix = {arXiv},
  primaryClass  = {hep-ph}
}

@article{GonzalezGarcia:2004wg,
  author    = {M.C. Gonzalez-Garcia and Michele Maltoni},
  title     = {Atmospheric Neutrino Oscillations and New Physics},
  journal   = {Phys. Rev. D},
  volume    = {70},
  pages     = {033010},
  year      = {2004},
  reportnumber = {YITP-SB-18-04},
  doi       = {10.1103/PhysRevD.70.033010},
  eprint    = {hep-ph/0404085},
  archivePrefix = {arXiv},
  primaryClass = {hep-ph}
}

@article{Diego:2008zu,
    author = "Diego, D. and Quiros, M.",
    title = "{Dirac Versus Majorana Neutrino Masses From a TeV Interval}",
    eprint = "0804.2838",
    archivePrefix = "arXiv",
    primaryClass = "hep-ph",
    reportNumber = "UAB-FT-643",
    doi = "10.1016/j.nuclphysb.2008.07.019",
    journal = "Nucl. Phys. B",
    volume = "805",
    pages = "148--167",
    year = "2008"
}

@article{Agashe:2000rw,
    author = "Agashe, Kaustubh and Wu, Guo-Hong",
    title = "{Remarks on models with singlet neutrino in large extra dimensions}",
    eprint = "hep-ph/0010117",
    archivePrefix = "arXiv",
    reportNumber = "OITS-696",
    doi = "10.1016/S0370-2693(00)01400-3",
    journal = "Phys. Lett. B",
    volume = "498",
    pages = "230--236",
    year = "2001"
}

@article{Barbieri:2000mg,
    author = "Barbieri, Riccardo and Creminelli, Paolo and Strumia, Alessandro",
    title = "{Neutrino oscillations from large extra dimensions}",
    eprint = "hep-ph/0002199",
    archivePrefix = "arXiv",
    reportNumber = "IFUP-TH-2000-00, SNS-PH-00-04",
    doi = "10.1016/S0550-3213(00)00348-5",
    journal = "Nucl. Phys. B",
    volume = "585",
    pages = "28--44",
    year = "2000"
}

@article{Mohapatra:2000wn,
    author = "Mohapatra, R. N. and Perez-Lorenzana, Abdel",
    title = "{Three flavor neutrino oscillations in models with large extra dimensions}",
    eprint = "hep-ph/0006278",
    archivePrefix = "arXiv",
    reportNumber = "UMD-PP-00-083",
    doi = "10.1016/S0550-3213(00)00634-9",
    journal = "Nucl. Phys. B",
    volume = "593",
    pages = "451--470",
    year = "2001"
}

@article{Esmaili:2014esa,
    author = "Esmaili, Arman and Peres, O. L. G. and Tabrizi, Zahra",
    title = "{Probing Large Extra Dimensions With IceCube}",
    eprint = "1409.3502",
    archivePrefix = "arXiv",
    primaryClass = "hep-ph",
    doi = "10.1088/1475-7516/2014/12/002",
    journal = "JCAP",
    volume = "12",
    pages = "002",
    year = "2014"
}

@article{DiIura:2014csa,
    author = "Di Iura, A. and Girardi, I. and Meloni, D.",
    title = "{Probing new physics scenarios in accelerator and reactor neutrino experiments}",
    eprint = "1411.5330",
    archivePrefix = "arXiv",
    primaryClass = "hep-ph",
    reportNumber = "RM3-TH-14-17, SISSA-62-2014-FISI",
    doi = "10.1088/0954-3899/42/6/065003",
    journal = "J. Phys. G",
    volume = "42",
    pages = "065003",
    year = "2015"
}

@article{Berryman:2016szd,
    author = "Berryman, Jeffrey M. and de Gouv{\^e}a, Andr{\'e} and Kelly, Kevin J. and Peres, O. L. G. and Tabrizi, Zahra",
    title = "{Large, Extra Dimensions at the Deep Underground Neutrino Experiment}",
    eprint = "1603.00018",
    archivePrefix = "arXiv",
    primaryClass = "hep-ph",
    doi = "10.1103/PhysRevD.94.033006",
    journal = "Phys. Rev. D",
    volume = "94",
    number = "3",
    pages = "033006",
    year = "2016"
}

@article{Basto-Gonzalez:2021aus,
    author = "Basto-Gonzalez, V. S. and Forero, D. V. and Giunti, C. and Quiroga, A. A. and Ternes, C. A.",
    title = "{Short-baseline oscillation scenarios at JUNO and TAO}",
    eprint = "2112.00379",
    archivePrefix = "arXiv",
    primaryClass = "hep-ph",
    doi = "10.1103/PhysRevD.105.075023",
    journal = "Phys. Rev. D",
    volume = "105",
    number = "7",
    pages = "075023",
    year = "2022"
}

@article{Khan:2022bcl,
    author = "Khan, Amir N.",
    title = "{Extra dimensions with light and heavy neutral leptons: an application to CE{\ensuremath{\nu}}NS}",
    eprint = "2208.09584",
    archivePrefix = "arXiv",
    primaryClass = "hep-ph",
    doi = "10.1007/JHEP01(2023)052",
    journal = "JHEP",
    volume = "01",
    pages = "052",
    year = "2023"
}

@misc{repolink,
  howpublished = {\url{https://github.com/WeiMXi/LED.git}}
}

@article{Lin:2023xyk,
    author = "Lin, Hai-Xing and Tang, Jian and Vihonen, Sampsa",
    title = "{Ultralight dark matter in neutrino oscillations to accommodate T2K and NO$\nu$A tension}",
    eprint = "2312.11704",
    archivePrefix = "arXiv",
    primaryClass = "hep-ph",
    month = "12",
    year = "2023"
}

@article{Abubakar_2026,
    author = "Abubakar, S. and others",
    collaboration = "NOvA",
    title = "{Precision Measurement of Neutrino Oscillation Parameters with 10 Years of Data from the NOvA Experiment}",
    eprint = "2509.04361",
    archivePrefix = "arXiv",
    primaryClass = "hep-ex",
    reportNumber = "FERMILAB-PUB-25-0619-PPD",
    doi = "10.1103/x53y-2b86",
    journal = "Phys. Rev. Lett.",
    volume = "136",
    number = "1",
    pages = "011802",
    year = "2026"
}

@article{Super-Kamiokande:2023ahc,
    author = "Wester, T. and others",
    collaboration = "Super-Kamiokande",
    title = "{Atmospheric neutrino oscillation analysis with neutron tagging and an expanded fiducial volume in Super-Kamiokande I{\textendash}V}",
    eprint = "2311.05105",
    archivePrefix = "arXiv",
    primaryClass = "hep-ex",
    doi = "10.1103/PhysRevD.109.072014",
    journal = "Phys. Rev. D",
    volume = "109",
    number = "7",
    pages = "072014",
    year = "2024"
}

@article{T2K:2024wfn,
    author = "Abe, K. and others",
    collaboration = "T2K, Super-Kamiokande",
    title = "{First Joint Oscillation Analysis of Super-Kamiokande Atmospheric and T2K Accelerator Neutrino Data}",
    eprint = "2405.12488",
    archivePrefix = "arXiv",
    primaryClass = "hep-ex",
    doi = "10.1103/PhysRevLett.134.011801",
    journal = "Phys. Rev. Lett.",
    volume = "134",
    number = "1",
    pages = "011801",
    year = "2025"
}

@article{IceCubeCollaboration:2024ssx,
    author = "Abbasi, R. and others",
    collaboration = "IceCube",
    title = "{Measurement of Atmospheric Neutrino Oscillation Parameters Using Convolutional Neural Networks with 9.3 Years of Data in IceCube DeepCore}",
    eprint = "2405.02163",
    archivePrefix = "arXiv",
    primaryClass = "hep-ex",
    doi = "10.1103/PhysRevLett.134.091801",
    journal = "Phys. Rev. Lett.",
    volume = "134",
    number = "9",
    pages = "091801",
    year = "2025"
}

\end{document}